\documentclass[11pt,a4paper]{article}
\usepackage{amsmath,amssymb,epsfig,a4,latexsym}
\usepackage{physics}
\usepackage{slashed}
\usepackage{jheppub}
\usepackage{color}
\usepackage{rotating}
\usepackage{graphicx}
\usepackage{caption}
\usepackage[caption=false]{subfig}

\newcommand{\Eqn}[1]{Eq.~(\ref{#1})}
\newcommand{\Eqns}[2]{Eqs.~(\ref{#1}) and (\ref{#2})}

\newcommand{\nc}{\newcommand}
\newcommand{\be }{\begin{equation}}   \nc{\ee }{\end{equation}}
\newcommand{\bea}{\begin{eqnarray}}   \nc{\eea}{\end{eqnarray}}
\newcommand{\baa}{\begin{array}}      \nc{\eaa}{\end{array}}
\newcommand{\bit}{\begin{itemize}}    \nc{\eit}{\end{itemize}}
\newcommand{\ben}{\begin{enumerate}}  \nc{\een}{\end{enumerate}}

\newcommand{\GeV}{\,\text{GeV}}
\newcommand{\Br}{\text{Br}}

\def\ben{\begin{enumerate}}
\def\een{\end{enumerate}}
\def\bit{\begin{itemize}}
\def\eit{\end{itemize}}
\def\beq{\begin{equation}}
\def\eeq{\end{equation}}
\def\bea{\begin{eqnarray}}
\def\eea{\end{eqnarray}}
\def \lsim{\mathrel{\vcenter
     {\hbox{$<$}\nointerlineskip\hbox{$\sim$}}}}

\def\meg{\mu \to e \gamma}

\begin{document}
\thispagestyle{empty}
\begin{flushright}
PSI-PR-17-01\\
ZU-TH 01/17\\
\today\\
\end{flushright}
\vspace{3em}
\begin{center}
{\Large\bf Renormalisation-group improved analysis of $\mu\to e$ processes\\[5pt]
in a systematic effective-field-theory approach} 
\\ 
\vspace{3em}
{\sc   A.~Crivellin$^{a,}$\footnote{e-mail: Andreas.Crivellin@cern.ch},
	S.~Davidson$^{b,}$\footnote{e-mail: S.Davidson@ipnl.in2p3.fr}, 
  G.~M.~Pruna$^{a,}$\footnote{e-mail: Giovanni-Marco.Pruna@psi.ch} and
	A.~Signer$^{a,c,}$\footnote{e-mail: Adrian.Signer@psi.ch}
}\\[2em]
{\sl ${}^a$ Paul Scherrer Institut,\\
CH-5232 Villigen PSI, Switzerland \\
\vspace{0.3cm}
${}^b$ IPNL, CNRS/IN2P3, 4 rue E. Fermi,\\
69622 Villeurbanne cedex, France;\\ 
Universit\'e Lyon 1, Villeurbanne;
Universit\'e de Lyon, F-69622, Lyon, France\\
\vspace{0.3cm}
${}^c$ Physik-Institut, Universit\"at Z\"urich, \\
Winterthurerstrasse 190,
CH-8057 Z\"urich, Switzerland}
\setcounter{footnote}{0}
\end{center}
\vspace{2ex}
\begin{abstract}
{}In this article, a complete analysis of the three muonic
lepton-flavour violating processes $\mu\to e \gamma$, $\mu\to 3e$ and
coherent nuclear $\mu\to e$ conversion is performed in the framework
of an effective theory with dimension six operators defined below the
electroweak symmetry breaking scale $m_W$.  The renormalisation-group
evolution of the Wilson coefficients between $m_W$ and the
experimental scale is fully taken into account at the leading order in
QCD and QED, and explicit analytic and numerical evolution matrices
are given. As a result, muonic decay and conversion rates are
interpreted as functions of the Wilson coefficients at any scale up to
$m_W$. Taking the experimental limits on these processes as input, the
phenomenology of the mixing effects is investigated. It is found that
a considerable set of Wilson coefficients unbounded in the simplistic
tree-level approach are instead severely constrained. In addition,
correlations among operators are studied both in the light of current
data and future experimental prospects.

\end{abstract}
 
\newpage
\setcounter{page}{1}

 \noindent\hrulefill
 \tableofcontents
 \noindent\hrulefill

\setcounter{footnote}{0}
\renewcommand{\thefootnote}{\arabic{footnote}}

\section{Introduction \label{sec:intro}}

Processes involving charged lepton flavour violation (LFV) are very
promising places to search for physics beyond the Standard Model
(BSM).  In the SM with massless neutrinos these processes are
forbidden, and even including neutrino masses and mixing these
processes are suppressed by ratios $m_\nu^4/m_W^4$.  Thus, they are by
far too small to be measured in any foreseeable experiment. Therefore,
any observation of charged LFV would directly prove the existence of
BSM physics.  In fact, in many BSM scenarios, such processes can have
sizeable decay rates which are naturally in the reach of forthcoming
experiments.

Among the charged LFV processes, the category of muonic transitions is
particularly interesting, providing three processes with complementary
sensitivities to new physics (NP) and stringently constrained by
experiment: the current limits from the
MEG~\cite{TheMEG:2016wtm,Adam:2013mnn} and
SINDRUM~\cite{Bellgardt:1987du,Bertl:2006up} collaborations are
\begin{eqnarray}
	&&{\rm Br}\left( \mu^+ \to e^+ \gamma \right)\leq4.2 \times10^{-13}\,,\\
	&&{\rm Br}\left(\mu^+\to e^+e^-e^+ \right)\leq1.0 \times10^{-12}\,,\\
	&&{\rm Br}_{\mu \to e}^{\rm Au}  \equiv
    \frac{\Gamma(\mu^-\,{\rm Au}\to e^-\,{\rm Au})}{\Gamma^{\rm capt}_{{\rm Au}}}
     \leq 7 \times 10^{-13}\,.
\end{eqnarray}
The future experimental prospects for $\mu\to e$ transitions are also
promising.  With the upgrade to MEG~II~\cite{Baldini:2013ke}, the
sensitivity of ${\rm Br}(\mu \to e \gamma)$ will reach $\sim 5\times 10^{-14}$.
Furthermore, Mu3e will improve the sensitivity on $\mu\to 3e$ by up to
$4$ orders of magnitude~\cite{Blondel:2013ia}. Concerning $\mu\to e$
conversion in nuclei, the DeeMe experiment aims at an accuracy of
$10^{-14}$~\cite{Aoki:2012zza}, while Mu2e at FNAL and COMET at
J-PARC~\cite{Carey:2008zz,Kutschke:2011ux,Cui:2009zz} aim to improve
the sensitivity to the conversion rate by four orders of magnitude
compared to SINDRUM~II. Moreover, the PRISM/PRIME
project~\cite{Barlow:2011zza} aims to reach the remarkable limit of
Br$(\mu \to e) \lsim 10^{-18}$.

LFV processes have been studied in detail in many specific extensions
of the SM (for a recent review see for example \cite{Lindner:2016bgg} and references therein). However, in this article an effective-field-theory
description of NP interactions is adopted. In this context, Kuno and
Okada~\cite{Kuno:1999jp} reviewed $\mu \to e$ flavour-changing
processes and experiments, and the operator basis required to
parameterise them. 

Constraints on Wilson coefficients, usually at the scale of the
experiments, have been compiled for LFV 2-lepton-2-quark
operators~\cite{Davidson:1993qk,Carpentier:2010ue} with a
tau-lepton~\cite{Dassinger:2007ru,Celis:2014asa,Celis:2014roa}, and 4-lepton
operators~\cite{Raidal:2008jk}. In the quark flavour
sector, a long-time effort allowed to establish the QCD running of
Wilson coefficients (see~\cite{Buras:1998raa} and references therein).

However, given that the electromagnetic coupling is much smaller than
the strong one, $\alpha_{e} \ll \alpha_s$, QED running is often
neglected for the leptons. Czarnecki and
Jankowski~\cite{Czarnecki:2001vf} proved that the self-renormalisation
of the $\meg$ dipole reduces the coefficient at low energy. Also in
the case of $(g-2)_\mu$ (lepton-flavour conserving effective
interactions), the QED running and mixing among a basis of SM-induced
operators has been included~\cite{Czarnecki:2002nt}.
More recently, the one-loop running of an $SU(2)$-invariant operator
basis~\cite{Buchmuller:1985jz,Grzadkowski:2010es} was presented in
\cite{Jenkins:2013wua, Alonso:2013hga} and loop effects in the
$SU(2)$-invariant theory were
calculated~\cite{Crivellin:2013hpa,Pruna:2014asa}\footnote{For an
  analogous analysis in the quark sector see for
  example~\cite{Alonso:2014csa, Alonso:2015sja,
    Calibbi:2015kma,Aebischer:2015fzz}.}. The results of
\cite{Jenkins:2013wua,Alonso:2013hga,Pruna:2014asa} were used in
\cite{Davidson:2016edt}, to translate the current $\mu\to
e\gamma$ bound from the experimental scale to the new-physics scale,
using QED$\times$QCD invariant operators below $m_W$, and the
$SU(2)$-invariant operators above\footnote{Restricting to the
  $SU(2)$-invariant operators below $m_W$ fails to consider relevant
  contributions, as shown in~\cite{Davidson:2016edt, Davidson:2016utf}.}.

The purpose of this article is to give the renormalisation-group
evolution (RGE) between the electroweak-symmetry-breaking (EWSB) scale
and the scale at which the experiments are performed. We consider the
Wilson coefficients that are relevant for $\mu\to e \gamma$, $\mu\to
3e$ and coherent $\mu\to e$ conversion and include the lowest
non-vanishing order in QED and QCD ($\equiv$ leading order). Our
analysis extends \cite{Davidson:2016edt} by including the one-loop RGE
of vector operators, and the two-loop mixing of vectors to the
dipoles, as well as the inclusion of $\mu\to e $ conversion and
$\mu\to 3e$ in the phenomenological analysis. In addition, we include
the dimension-seven lepton-gluon operator that is relevant in $\mu\to
e$ conversion~\cite{Cirigliano:2009bz}.

The outline of the paper is as follows: Section~\ref{sec:2} introduces
the QCD$\times$QED invariant operators of our effective Lagrangian.
The essential formulas for the transition rates of our three processes
in terms of the Wilson coefficients of the operators are recalled in
Section~\ref{sec:processes}.  Then, the RGE of these Wilson
coefficients is discussed in Section~\ref{sec:running} with the
analytic formulas collected in Appendix~\ref{app:adim}.  These results
are combined in Section~\ref{sec:pheno} to obtain limits on various
Wilson coefficients and discuss the complementarity of the three
processes.

\section{Low-energy Lagrangian}
\label{sec:2}

Following the Appelquist-Carazzone theorem~\cite{Appelquist:1974tg},
we consider an effective Lagrangian that is valid below some scale
$\Lambda$ with $m_{W} \ge \Lambda \gg m_b$. Therefore, it consists of
all operators that are invariant under $U(1)_{\rm QED} \times
SU(3)_{\rm QCD}$ and contain the fermion fields
$f\in\{u,d,c,s,b,e,\mu,\tau\}$, as well as the QED and QCD gauge
fields.  In addition to the dimension-four QED and QCD Lagrangians, it
contains higher-dimensional operators multiplied by dimensionless
Wilson coefficients $C$. Having $\mu\to e$ transitions in mind, we
restrict ourselves to operators that induce such transitions and are
flavour diagonal with respect to the other fields.  Concretely, our
Lagrangian reads
\begin{align}
\mathcal{L}_{\rm eff}&=\mathcal{L}_{\rm QED} + \mathcal{L}_{\rm QCD} 
\nonumber \\[5pt]
&+\frac{1}{\Lambda^2}\bigg\{C_L^DO_L^D 
+ \sum\limits_{f = q,\ell } {\left( 
      {C_{ff}^{V\;LL}O_{ff}^{V\;LL} + C_{ff}^{V\;LR}O_{ff}^{V\;LR} 
     + C_{ff}^{S\;LL}O_{ff}^{S\;LL}} \right)}\nonumber \\
& \qquad
+ \sum\limits_{h = q,\tau } {\left( 
       {C_{hh}^{T\;LL}O_{hh}^{T\;LL} + C_{hh}^{S\;LR\;}O_{hh}^{S\;LR\;}} \right)}+
C_{gg}^LO_{gg}^L  + L \leftrightarrow R\bigg\} + \mbox{h.c.},
\label{Leff}
\end{align}
with the explicit form of the operators given by
\begin{align}
\label{eq:magnetic}
O_L^{D} &= e\, m_\mu\left( \bar e{\sigma ^{\mu \nu }}{P_L}\mu\right) {F_{\mu \nu }},
\\[3pt]
O_{ff}^{V\;LL} &= \left(\bar e{\gamma ^\mu }{P_L}\mu\right) 
\left( \bar f{\gamma _\mu }{P_L}f\right),
\\[3pt]
O_{ff}^{V\;LR} &= \left(\bar e{\gamma ^\mu }{P_L}\mu\right) 
\left( \bar f{\gamma _\mu }{P_R}f\right),
\\[3pt]
O_{ff}^{S\;LL} &= \left(\bar e{P_L}\mu\right) \left( \bar f{P_L}f\right),
\\[3pt]
O_{hh}^{S\;LR} &= \left(\bar e{P_L}\mu\right) \left( \bar h{P_R}h\right),
\\[3pt]
O_{hh}^{T\;LL} &= \left(\bar e{\sigma _{\mu \nu }}{P_L}\mu\right) 
\left( \bar h{\sigma ^{\mu \nu }}{P_L}h\right),
\\[3pt]
O_{gg}^L &= \alpha_s\, m_\mu G_F
\left( \bar e P_L\mu  \right)G^a_{\mu \nu }G_a^{\mu \nu },
\label{Ogg}
\end{align}
where $P_{L/R}=\left(\mathbb{I}\mp \gamma^5\right)/2$.  The
field-strength tensors for photons and gluons are denoted by
$F^{\mu\nu}$ and $G^{\mu\nu}_a$, respectively. Regarding the matter
fields, $f$ represents any fermion below the decoupling scale $m_W$,
and $h\in\{u,d,c,s,b,\tau\}$ is restricted to be either a quark
(hadron) or the $\tau$ (heavy lepton). Note that for $f\in \{e,\mu\}$,
the tensor $O_{ff}^{T\;LL}$, $O_{ff}^{T\;RR}$ and scalar
$O_{ff}^{S\;LR}$, $O_{ff}^{S\;RL}$ operators can be reduced by Fierz
transformations to other operators already present in
$\mathcal{L}_{\rm eff}$. Therefore, this Lagrangian does not contain
redundant operators and constitutes a minimal basis.

Concerning the dipole operators $O_L^{D}$ and $O_R^{D}$, the fields
only add up to dimension~5. Thus, they cannot mix into the
four-fermion operators (due to renormalisability arguments), although
the latter can mix into the dipole operators. Since the dipole
operators flip chirality we have defined them by including a
factor of $m_\mu$, enhancing their dimensionality up to 6. Also,
the prefactor $e$ included in $O_L^{D}$ and $O_R^{D}$ is introduced
for convenience (when considering the RGE).

Following \cite{Cirigliano:2009bz}, the only dimension~7 operators
that we include in Eq.~(\ref{Ogg}) are $O_{gg}^L$ and
$O_{gg}^R$. These operators are phenomenologically relevant because,
as shown in \cite{Shifman:1978zn}, they encode the effects of scalar
operators with heavy quarks ({\it i.e.} $c,b$) below the heavy-quark
mass scale $m_Q$, after the matching has been performed. In practical
terms, these operators are suppressed by $1/(\Lambda^2 m_Q)$ rather
than $1/(\Lambda^3)$.  The normalisation of these operators has been
chosen such that their Wilson coefficients do not run under QCD at one
loop and the factor $G_F$ is included to resize the dimensionality
down to 6.

In the scenario where BSM physics is realised at a scale $\Lambda <
m_W$, this NP directly gives rise to the higher-dimensional operators
in $\mathcal{L}_{\rm eff}$.  If BSM physics is beyond the EWSB scale,
as it is usually assumed, $SU(2)$ invariant higher-dimensional
operators are first generated in the Standard Model Effective Field
Theory (SMEFT).  Then, the higher-dimensional operators in
$\mathcal{L}_{\rm eff}$ stem from the matching of the SMEFT to our
theory, as performed at tree level in \cite{Davidson:2016edt}.  It is
clear that $\mathcal{L}_{\rm eff}$ cannot be a valid description of
nature above the EWSB scale, as it does not respect the $SU(2)$
symmetry.

However, we stress the fact that if NP is realised not too far above
the EWSB scale, then a reduced hierarchy between the NP and EWSB
scales would not allow for potential large logarithms from the SMEFT
RGE, while the hierarchy between the EWSB scale (our matching scale)
and the muon/nuclear scale is always sufficiently large to give rise
to important effects.

\section{LFV muon decays \label{sec:processes}}

In this section, the expressions for the processes $\mu^+ \to
e^+\gamma$, $\mu^+\to e^+ e^- e^+$, and coherent muon-to-electron
conversion in muonic atoms $\mu^- N\to e^- N$ in terms of the
coefficients $C_i$ of the Lagrangian~\Eqn{Leff} will be presented. We
restrict ourselves to the tree-level expressions. Loop effects are
only included through the RGE of the Wilson coefficients. Thus, for
the expressions that follow the Wilson coefficients are understood to
be evaluated at the scale of the process.

\begin{boldmath}
\subsection{$\mu\to e \gamma$}
\end{boldmath}
Currently, $\mu^+\to e^+\gamma$ is the LFV muonic process with the
most stringent experimental bound~\cite{TheMEG:2016wtm}. At the
tree level, the Lagrangian in \Eqn{Leff} results in the branching ratio
\begin{align}
\label{muegBR}
{\rm{Br}}\left( {\mu  \to e\gamma } \right)  =  
\frac{\alpha_e m_\mu ^5}{{\Lambda^4 \Gamma _\mu }}\left( 
{{{\left| {C^{D}_L} \right|}^2} 
+ {{\left| {C^{D}_R} \right|}^2}} \right)\,,
\end{align}
where $\Gamma _\mu$ is the width of the muon. The scale of the process
is $\mu=m_\mu$. Operators other than the dipole will enter this
process only through the RGE.

\begin{boldmath}
\subsection{$\mu\to eee$}
\end{boldmath}

The bounds on $\mu\to 3e$ are already very precise, but Mu3e will
significantly improve them in the coming years~\cite{Blondel:2013ia}.
The branching ratio expressed in terms of Wilson coefficients
$C_i(m_\mu)$ reads
\begin{align}
{\rm Br}(\mu  \to 3e) &=
\frac{\alpha_e^2 m_\mu^5 }{12 \pi \Lambda^4 \Gamma _\mu}
\left(\left|C^{D}_{L}\right|^2+\left|C^{D}_{R}\right|^2\right)
\left(8 \log\left[\frac{m_\mu}{m_e}\right]-11\right)+ X_\gamma
\label{br:meee} \\ 
&+\frac{m_\mu^5}{3  (16\pi)^3 \Lambda^4 \Gamma _\mu}
\bigg(\, \left|C_{ee}^{S\;LL}\right|^2 + 
16 \left|C_{ee}^{V\;LL}\right|^2 + 8 \left|C_{ee}^{V\;LR}\right|^2  \nonumber\\
& \phantom{+\frac{m_\mu^5}{3  (16\pi)^3 \Lambda^4 \Gamma _\mu}} 
+\
\left|C^{S\;RR}_{ee}\right|^2 +
16 \left|C_{ee}^{V\;RR}\right|^2 + 8 \left|C_{ee}^{V\;RL}\right|^2\bigg) \, ,
\nonumber
\end{align}
where the interference term with the dipole operator is given by
\begin{align}
X_\gamma ^{}{\rm{ }} =
-\frac{\alpha_e m_\mu^5 }{3 (4\pi)^2 \Lambda^4 \Gamma _\mu}
  (\Re[C^{D}_{L}
  \left(C_{ee}^{V\;RL}+2C_{ee}^{V\;RR}\right)^*]+\Re[C^{D}_{R}
  \left(2C_{ee}^{V\;LL}+C_{ee}^{V\;LR}\right)^*])\, .
\end{align}
Notice that interference between the four-lepton operators can be
neglected because it is suppressed by powers of $m_e$. So in contrast
to $\mu\to e \gamma$ and coherent $\mu\to e$ conversion, which only
impose two constraints, the upper bound on $\Br(\mu\to 3e)$ sets
independent constraints on several four-lepton operators.  Through the
RGEs, this process is also sensitive to operators involving quarks or
other leptons.

\begin{boldmath}
\subsection{$\mu\to e$ conversion in nuclei}
\label{sec:conversion}
\end{boldmath}

The coherent LFV muonic transition in nuclei is produced by the
dipole, the vector, and the scalar operators already at the tree
level~\cite{Kitano:2002mt}.  Once relativistic and finite nuclear size
effects are taken into account for heavy
nuclei~\cite{Shanker:1979ap,Czarnecki:1998iz, Kitano:2002mt}, the
transition amplitudes for the three classes of operators exhibit
different sensitivities to the atomic number $Z$. Hence, in principle
different target atoms provide different limits on the coefficients of
the involved class of operators~\cite{Cirigliano:2009bz}.  Whether it is feasible to perform
measurements for many different elements remains to be seen.  The
SINDRUM collaboration has presented limits for gold, titanium and
lead~\cite{Bertl:2006up,Dohmen:1993mp, Honecker:1996zf}, but the
upcoming experiments mostly concentrate on aluminium.

For this process, the Lagrangian $\mathcal{L}_{\rm eff}$ as given in
\Eqn{Leff} is not directly suitable. Instead, a Lagrangian at the
nucleon level containing proton and neutron fields is required. This
Lagrangian is obtained in two steps. First, heavy quarks are
integrated out. This results in a redefinition of the Wilson
coefficient of the gluonic operator~\cite{Shifman:1978zn}
\begin{align}
C_{gg}^L &\to \tilde{C}_{gg}^L = C_{gg}^L - \frac{1}{12 \pi}
\sum_{q=c,b} \frac{C_{qq}^{S\;LL}+C_{qq}^{S\;LR}}{G_F\, m_\mu m_q}
\label{anomaly}
\end{align}
with an analogous equation for $C_{gg}^R$. Second, the resulting
Lagrangian is matched at a scale of $\mu_n=1$~GeV to an effective
Lagrangian at the nucleon level.  Following~\cite{Cirigliano:2009bz}
the transition rate $\Gamma_{\mu\to e}^N = \Gamma(\mu^- N \to e^- N)$
can then be written as
\begin{align}
\Gamma_{\mu\to e}^N &= \frac{m_{\mu}^{5}}{4\Lambda^4}
\left| e\,  C^{D}_{L} \; D_N + 
4\left(
G_F m_\mu m_p \tilde{C}_{(p)}^{SL}  S^{(p)}_N
+   \tilde{C}_{(p)}^{VR} \; V^{(p)}_N
+ p \to n \right) \right|^2 +L\leftrightarrow R,
\label{Gconv}
\end{align}
where $p$ and $n$ denote the proton and the neutron, respectively. The
effective couplings in \Eqn{Gconv} can be expressed in terms of our
Wilson coefficients as
\begin{align}
\tilde{C}_{(p/n)}^{VR} &= \sum_{q=u,d,s} 
\left(C_{qq}^{V\;RL}+C_{qq}^{V\;RR}\right) \; f^{(q)}_{Vp/n} \, , 
\label{tildeCVR} \\
\tilde{C}_{(p/n)}^{SL} &=  \sum_{q=u,d,s} 
\frac{\left(C_{qq}^{S\;LL}+C_{qq}^{S\;LR}\right)}{m_\mu m_q G_F} 
\; f^{(q)}_{Sp/n}  \ + \ \tilde{C}_{gg}^L \, f_{Gp/n} 
\label{tildeCSL}
\end{align}
with analogous relations for $L\leftrightarrow R$. The Wilson
coefficients in \Eqns{tildeCVR}{tildeCSL} are to be evaluated at the
scale $\mu_n$.

The nucleon form factors for vector operators are known from the
vector-current conservation, \emph{i.e.} $f_{Vp}^{(u)}=2$,
$f_{Vn}^{(u)}=1$, $f_{Vp}^{(d)}=1$, $f_{Vn}^{(d)}=2$,
$f_{Vp}^{(s)}=0$, $f_{Vn}^{(s)}=0$. Hence, the sum in \Eqn{tildeCVR}
is in fact only over $q=u ,d$. The calculation of the scalar form
factors is more involved. Following~\cite{Crivellin:2014cta}, the values of the up- and
down-quark scalar couplings $f_{S p/n}^{(u/d)}$ from
\cite{Hoferichter:2015dsa} (based on the two-flavour chiral
perturbation theory framework of~\cite{Crivellin:2013ipa}) are used.
The values of the $s$-quark scalar couplings $f_{S p/n}^{(s)}$ are
borrowed from a lattice calculation~\cite{Junnarkar:2013ac}\footnote{See~\cite{Alarcon:2011zs,Alarcon:2012nr} for analogous results obtained from a computation in an effective-field theoretical framework.}. In
summary, we use
\begin{align}
f_{Sp}^{(u)}&=(20.8\pm 1.5) \times 10^{-3},\qquad
f_{Sn}^{(u)}=(18.9\pm 1.4) \times 10^{-3},
\nonumber \\
f_{Sp}^{(d)}&=(41.1\pm 2.8) \times 10^{-3},\qquad
f_{Sn}^{(d)}=(45.1\pm 2.7) \times 10^{-3},
\nonumber \\
f_{Sp}^{(s)}&=f_{Sn}^{(s)}=(53\pm 27)\times 10^{-3}.
\end{align}
The form factor for the gluonic operator can be obtained from a sum
rule. In our normalisation we get
\begin{align}
f_{Gp/n}&=-\frac{8\pi}{9}\Big(1-\sum_{q=u,d,s} f^{(q)}_{Sp/n}\Big) \, .
\end{align}
The quantities $D_N$, $S^{(p/n)}_N$, and $V^{(p/n)}_N$ in \Eqn{Gconv}
are related to the overlap integrals~\cite{Czarnecki:1998iz} between
the lepton wave functions and the nucleon densities.  They depend on
the nature of the target $N$ and we use the numerical
values~\cite{Kitano:2002mt}
\begin{align}
 D_{\rm Au}&=0.189, &S^{(p)}_{\rm Au}&=0.0614, &V^{(p)}_{\rm Au}&=0.0974,
&S^{(n)}_{\rm Au}&=0.0918, &V^{(n)}_{\rm Au}&=0.146;  \nonumber \\
 D_{\rm Al}&=0.0362, &S^{(p)}_{\rm Al}&=0.0155, &V^{(p)}_{\rm Al}&=0.0161, 
&S^{(n)}_{\rm Al}&=0.0167, &V^{(n)}_{\rm Al}&=0.0173;
\end{align}
for gold and aluminium, 
respectively. The branching ratio is defined as the transition rate,
\Eqn{Gconv}, divided by the capture rate. For the latter we use
\begin{align}
\Gamma^{\rm capt}_{{\rm Au}}&=8.7\times 10^{-15}~{\rm MeV},&
\Gamma^{\rm capt}_{{\rm Al}}&=4.6\times 10^{-16}~{\rm MeV},
\end{align}
taken from~\cite{Suzuki:1987jf}.

\section{Renormalisation-group evolution \label{sec:running}}

The operators present in $\mathcal{L}_{\rm eff}$, \Eqn{Leff}, will
give rise to $\mu\to e$ transitions.  Thus, experimental constraints
from these processes can be translated into bounds on the coefficients
of various operators.  However, in the first instance this procedure
results in bounds on $C(m_\mu)$ or $C(\mu_n)$, the Wilson coefficients
evaluated at the low scale.  In order to gain more insight into
possible NP scenarios, it is preferable to extract limits on
$C(\Lambda)$, {\emph{i.e.}} on the Wilson coefficients at the high
scale.  If $\Lambda < m_W$, this will be directly at the NP scale. If
the BSM scale is higher than the EWSB scale, our theory will have to
be matched to the SMEFT and, in a second step, the RGE within the
SMEFT~\cite{Jenkins:2013wua, Alonso:2013hga} will have to be carried
out from $m_W$ to the scale, where the BSM theory is to be matched to
the SMEFT.

Under the RGE, the various operators in $\mathcal{L}_{\rm eff}$ mix
among each other.  In our analysis, we take into account at least the
one-loop anomalous dimensions for all the operators.  Since the dipole
operator plays a prominent role in all $\mu\to e$ transitions, we also
consider those two-loop effects of direct mixing into $C^D_L$ and
$C^D_R$ that are ``leading'', {\emph{i.e.} those with vanishing
  one-loop contribution.

At one loop, the structure of the anomalous dimension matrix splits
into two blocks, the vector operators and all the other operators. The
dipole operator $Q^D_L$ receives contributions from
itself~\cite{Czarnecki:2001vf, Pruna:2014asa}, $Q_{ee}^{S\;LL}$,
$Q_{\mu\mu}^{S\;LL}$ and $Q_{hh}^{T\;LL}$~\cite{Cirigliano:2009bz,
  Davidson:2016edt}, and the scalar and tensor operators also mix
among themselves.  In addition the dimension~7 gluon operators mixes
into the quark scalar operators.  On the other side, the vector
operators mix among themselves~\cite{Buras:1991jm}, but are not
connected to the remaining operators.

Many of the anomalous dimensions we need can be found in the
literature. However, we have performed an independent computation of
all one-loop anomalous dimensions in our basis. In order to perform
such a calculation in an automated way, several openly available tools
were used: the described model was implemented in {\tt FeynRules~v2.3}~\cite{Alloul:2013bka} to obtain consistent Feynman
rules\footnote{The model file is available upon request.}; the
FeynArts interface of FeynRules was exploited to produce a model file
for the {\tt FeynArts~v3.9}~\cite{Hahn:2000kx} and {\tt FormCalc~v9.2}~\cite{Hahn:1998yk,Hahn:2016ebn} packages; the combined
packages FeynArts/FormCalc were employed to generate non-integrated
amplitudes to be elaborated afterwards with the symbolic manipulation
system {\tt Form~v4.1}~\cite{Kuipers:2012rf}.

At two loops, also the vector operators mix into dipole
operators. Since this mixing is not induced through one-loop diagrams,
then the two-loop contributions ${\cal O}(\alpha_e)$ are the leading
order effect. In general, the corresponding anomalous dimensions are
regularisation-scheme dependent. This issue has received a lot of
attention in the context of $b\to s \gamma$
transitions~\cite{Misiak:1993es, Ciuchini:1993fk}. For a physical
quantity, this scheme dependence is cancelled by the corresponding
scheme dependence of the one-loop finite contributions.  In fact, this
contribution is of the same order in the coupling, namely ${\cal
  O}(\alpha_e)$. Here, we present our two-loop anomalous dimensions for
vector operators in the 't~Hooft-Veltman (HV) scheme, where the finite
one-loop contribution vanishes. They have been extracted from
\cite{Ciuchini:1993fk} and are given in Appendix~\ref{app:adim}. This
can also be done for those scalar operators ($O_{hh}^{S\;LR}$ and
$O_{hh}^{S\;RL}$) that can be interchanged with vector operators by Fierz transformations.
We remark that our approach, while phenomenologically useful, it is not completely self-consistent and should not be understood as a replacement of a two-loop precision calculation, but rather as a qualitative indication of a leading order effect.

Assembling the Wilson coefficients in a vector $\vec{C}(\mu)$, the
renormalisation-group running can be written in matrix form as 
\bea
\mu \dv{}{\mu} \vec{C}(\mu) &=& 
\frac{1}{4\pi}\, {\Gamma}^0 \vec{C} +
\frac{\alpha_{s}}{4\pi}\, {\Gamma}^{s} \vec{C} +  
\frac{\alpha_{e}}{4\pi}\, {\Gamma}^e \vec{C}=\nonumber \\
&=&{\frac{8\pi^2}{e^2}\Gamma_{ST}\vec{C}+
\frac{36\pi^2}{e^2}\Gamma_{V}\vec{C}+
96\pi^2\Gamma_{VD}\vec{C}} \, ,
\label{RGE}
\eea 
where ${\Gamma}^i$ are the transpose of the anomalous dimension
matrices. {The analytic results for the ${\Gamma}_{ST}$ (scalar-tensor mixing) are shown in Table~\ref{Scalar}, and for the ${\Gamma}_{V}$ (vector mixing) and ${\Gamma}_{VD}$ (vector into dipole mixing) in the upper squared block and last row of Table~\ref{Vector}, respectively}. The complete expressions,
keeping small mass ratios and charge factors are collected in
Appendix~\ref{app:adim}.

\begin{table}[!th]
  \centering
  \resizebox{\textwidth}{!}{  
  \begin{tabular}{c||c|cc|cc|cc|c|c|c|c}
$
\begin{aligned}
  \frac{8\pi^2}{e^2}\mu\dv{}{\mu}
\end{aligned}
$
&
$C^{S\; LL}_{ee/\mu\mu}$ & $C^{S\; LL}_{\tau\tau}$ & $C^{T\; LL}_{\tau\tau}$ &
$C^{S\; LL}_{uu/cc}$ & $C^{T\; LL}_{uu/cc}$ & $C^{S\; LL}_{dd/ss/bb}$&
$C^{T\; LL}_{dd/ss/bb}$ & $C^{S\; LR}_{\tau\tau}$&$C^{S\; LR}_{uu/cc}$&
$C^{S\; LR}_{dd/ss/bb}$&$C^{D}_{L}$ \\[1.5ex]
\hline
\rule{0pt}{3.5ex}
$C^{S\; LL}_{ee/\mu\mu}$&
$6$ & $0$ & $0$ & $0$ & $0$ & $0$ & $0$ & $0$ & $0$ & $0$ & $0$ \\[1.5ex]
 \hline
\rule{0pt}{3.5ex}
$C^{S\; LL}_{\tau\tau}$&
$0$ & $-6$ & $-48$ & $0$ & $0$ & $0$ & $0$ & $0$ & $0$ & $0$ & $0$ \\[1.5ex]
$C^{T\; LL}_{\tau\tau}$&
$0$ & $-1$ & $2$ & $0$ & $0$ & $0$ & $0$ & $0$ & $0$ & $0$ & $0$ \\[1.5ex]
 \hline
\rule{0pt}{3.5ex}
$C^{S\; LL}_{uu/cc}$&
$0$ & $0$ & $0$ &
$
\begin{aligned}
-\frac{13}{3}-4\frac{g_s^2}{e^2}
\end{aligned}
$
& $32$ & $0$ & $0$ & $0$ & $0$ & $0$ & $0$ \\[1.5ex]

$C^{T\; LL}_{uu/cc}$ &
$0$ & $0$ & $0$ &
$\begin{aligned}
\frac{2}{3}
\end{aligned}$
&
$\begin{aligned}
\frac{13}{9}+4\frac{g_s^2}{e^2}
\end{aligned}$
&
$0$ & $0$ & $0$ & $0$ & $0$ & $0$ \\[1.5ex]
 \hline
\rule{0pt}{3.5ex}
$C^{S\; LL}_{dd/ss/bb}$ &
$0$ & $0$ & $0$ & $0$ & $0$ &
$\begin{aligned}
-\frac{10}{3}-4\frac{g_s^2}{e^2}
\end{aligned}$
&
$-16$ & $0$ & $0$ & $0$ & $0$\\[1.5ex]
$C^{T\; LL}_{dd/ss/bb}$ &
$0$ & $0$ & $0$ & $0$ & $0$ &
$\begin{aligned}
-\frac{1}{3}
\end{aligned}$
&
$\begin{aligned}
\frac{10}{9}+\frac{4}{3}\frac{g_s^2}{e^2}
\end{aligned}$&
$0$ & $0$ & $0$ & $0$ \\[1.5ex]
 \hline
\rule{0pt}{3.5ex}
$C^{S\; LR}_{\tau\tau}$ &
$0$ & $0$ & $0$ & $0$ & $0$ & $0$ & $0$ & $-6$ & $0$ & $0$ & $0$\\[1.5ex]
 \hline
\rule{0pt}{3.5ex}
$C^{S\; LR}_{uu/cc}$ &
$0$ & $0$ & $0$ & $0$ & $0$ & $0$ & $0$ & $0$ &
$\begin{aligned}
-\frac{13}{3}-4\frac{g_s^2}{e^2}
\end{aligned}$&
$0$ & $0$ \\[1.5ex]
 \hline
\rule{0pt}{3.5ex}
$C^{S\; LR}_{dd/ss/bb}$ &
$0$ & $0$ & $0$ & $0$ & $0$ & $0$ & $0$ & $0$ & $0$ &
$\begin{aligned}
-\frac{10}{3}-4\frac{g_s^2}{e^2}
\end{aligned}$&
$0$\\[1.5ex]
 \hline
\rule{0pt}{3.5ex}
$C^{D}_{L}$
&
$\begin{aligned}
\frac{m_{e/\mu}}{2e^2m_\mu}
\end{aligned}$&
$0$ &
$\begin{aligned}
-\frac{4m_\tau}{e^2m_\mu}
\end{aligned}$ &
$0$ &
$\begin{aligned}
\frac{8m_c}{e^2m_\mu}
\end{aligned}$ &
$0$ &
$\begin{aligned}
-\frac{4m_b}{e^2m_\mu}
\end{aligned}$&
$0$ & $0$ & $0$ & $-8$
\end{tabular}
}
  \caption{Entries of the transposed anomalous dimension matrix
  {${\Gamma}_{ST}$} for the scalar-tensor sector. }
  \label{Scalar}
\end{table}

\begin{table}[!th]
  \centering
  \resizebox{\textwidth}{!}{  
  \begin{tabular}{c||cccccc|cccccc|cccc}
$
\begin{aligned}
  \frac{36\pi^2}{e^2}\mu\dv{}{\mu}
\end{aligned}
$
&
$C^{V\; RR}_{ee}$ &$C^{V\; RR}_{\mu\mu}$ &$C^{V\; RR}_{\tau\tau}$ &
$C^{V\; RL}_{ee}$ & $C^{V\; RL}_{\mu\mu}$ &$C^{V\; RL}_{\tau\tau}$ &
$C^{V\; RR}_{dd}$ &$C^{V\; RR}_{ss}$ &$C^{V\; RR}_{bb}$ &
$C^{V\; RL}_{dd}$ &$C^{V\; RL}_{ss}$ &$C^{V\; RL}_{bb}$ &
$C^{V\; RR}_{uu}$ &$C^{V\; RR}_{cc}$ &$C^{V\; RL}_{uu}$ &$C^{V\; RL}_{cc}$  \\[1.5ex]
\hline
\rule{0pt}{3.5ex}
$C^{V\; RR}_{ee}$ & 33 & 6 & 3& 3& 3& 3&
3& 3& 3& 3&3&3&
-6&-6&-6&-6\\[1.5ex]
\rule{0pt}{3.5ex}
$C^{V\; RR}_{\mu\mu}$ & 6 & 33 & 3& 3& 3& 3&
3& 3& 3& 3&3&3&
-6&-6&-6&-6\\[1.5ex]
\rule{0pt}{3.5ex}
$C^{V\; RR}_{\tau\tau}$ & 6 & 6 & 30& 3& 3& 3&
3& 3& 3& 3&3&3&
-6&-6&-6&-6\\[1.5ex]
\rule{0pt}{3.5ex}
$C^{V\; RL}_{ee}$ & 6 & 6 & 3& -24& 3& 3&
3& 3& 3& 3&3&3&
-6&-6&-6&-6\\[1.5ex]
\rule{0pt}{3.5ex}
$C^{V\; RL}_{\mu\mu}$ & 6 & 6 & 3& 3& -24& 3&
3& 3& 3& 3&3&3&
-6&-6&-6&-6\\[1.5ex]
\rule{0pt}{3.5ex}
$C^{V\; RL}_{\tau\tau}$ & 6 & 6 & 3& 3& 3& -24&
3& 3& 3& 3&3&3&
-6&-6&-6&-6\\[1.5ex]
\hline
\rule{0pt}{3.5ex}
$C^{V\; RR}_{dd}$ & 2 & 2 & 1& 1& 1& 1&
10& 1& 1& 1&1&1&
-2&-2&-2&-2\\[1.5ex]
\rule{0pt}{3.5ex}
$C^{V\; RR}_{ss}$ & 2 & 2 & 1& 1& 1& 1&
1& 10& 1& 1&1&1&
-2&-2&-2&-2\\[1.5ex]
\rule{0pt}{3.5ex}
$C^{V\; RR}_{bb}$ & 2 & 2 & 1& 1& 1& 1&
1& 1& 10& 1&1&1&
-2&-2&-2&-2\\[1.5ex]
\rule{0pt}{3.5ex}
$C^{V\; RL}_{dd}$ & 2 & 2 & 1& 1& 1& 1&
1 & 1 & 1 & -8 & 1 & 1 &
-2&-2&-2&-2\\[1.5ex]
\rule{0pt}{3.5ex}
$C^{V\; RL}_{ss}$ & 2 & 2 & 1 & 1 & 1 & 1 &
1 & 1 & 1 & 1 & -8 & 1 &
-2 & -2 & -2 & -2 \\[1.5ex]
\rule{0pt}{3.5ex}
$C^{V\; RL}_{bb}$ & 2 & 2 & 1 & 1 & 1 & 1 &
1 & 1 & 1 & 1 & 1 & -8 &
-2 & -2 & -2 & -2 \\[1.5ex]
\hline
\rule{0pt}{3.5ex}
$C^{V\; RR}_{uu}$ & -4 & -4 & -2 & -2 & -2 & -2 &
-2 & -2 & -2 & -2 & -2 & -2 &
-14 & 4 & 4 & 4 \\[1.5ex]
\rule{0pt}{3.5ex}
$C^{V\; RR}_{cc}$ & -4 & -4 & -2 & -2 & -2 & -2 &
-2 & -2 & -2 & -2 & -2 & -2 &
4 & -14 & 4 & 4\\[1.5ex]
\rule{0pt}{3.5ex}
$C^{V\; RL}_{uu}$ & -4 & -4 & -2 & -2 & -2 & -2 &
-2 & -2 & -2 & -2 & -2 & -2 &
4 & 4 & 22 & 4\\[1.5ex]
\rule{0pt}{3.5ex}
$C^{V\; RL}_{cc}$ & -4 & -4 & -2 & -2 & -2 & -2 &
-2 & -2 & -2 & -2 & -2 & -2 &
4 & 4 & 4 & 22\\[1.5ex]
\hline
\rule{0pt}{3.5ex}
$96\pi^2C^{D}_{L}$ & 174 & 174 & 87 & -150 & -150 & -75 &
11 & 11 & 11 & -7 & -7 & -7 &
32 & 32 & -40 & -40
\end{tabular}
}
  \caption{Entries of the transposed anomalous dimension matrices
  {${\Gamma}_{V}$ (upper squared block) and ${\Gamma}_{VD}$ (last row)} for the vector operators. The last row determines the
  two-loop mixing of the vector operators into the dipole operator.}
  \label{Vector}
\end{table}

In the evolution, the operators involving bottom quarks, taus and
charm quarks are integrated down to the various thresholds. {Concerning the two-loop running of the dipole operator, we remark that the threshold effects result in vanishing one-loop contributions in the 't~Hooft-Veltman (HV) scheme.} Operators
involving light quarks are not meaningful any longer for scales below
$\mu_n=1\GeV$. At about this scale, our theory should be matched to a
nuclear effective theory (following the same approach that we adopted
for the $\mu\to e$ conversion in Section~\ref{sec:conversion}).
Concerning the RGE evolution, we also integrate out the light quark
operators at 1~GeV.

In order to illustrate the numerical importance of the various terms,
we express the Wilson coefficients at a low scale in terms of the
Wilson coefficients at $\mu=m_W$. To this end, we decompose the vector
$\vec{C} = (C^D_L, \vec{C}_l^{ST}, \vec{C}_q^{ST}, \vec{C}^V,
C_{gg}^L)$, where
\begin{align}
\vec{C}_l^{ST}= \big(&
C_{ee}^{S\;LL}, C_{\mu\mu}^{S\;LL},
C_{\tau\tau}^{S\;LL}, C_{\tau\tau}^{T\;LL}, C_{\tau\tau}^{S\;LR}\big),
 \nonumber
\\[5pt]
\vec{C}_q^{ST}= \big(&
C_{dd}^{S\;LL}, C_{dd}^{T\;LL}, C_{dd}^{S\;LR}, 
C_{ss}^{S\;LL}, C_{ss}^{T\;LL}, C_{ss}^{S\;LR}, 
C_{bb}^{S\;LL}, C_{bb}^{T\;LL}, C_{bb}^{S\;LR}, 
\nonumber \\
& 
C_{uu}^{S\;LL}, C_{uu}^{T\;LL}, C_{uu}^{S\;LR}, 
C_{cc}^{S\;LL}, C_{cc}^{T\;LL}, C_{cc}^{S\;LR} \big),
\\[5pt]
\vec{C}^V = \big(&
C_{ee}^{V\;RR}, C_{ee}^{V\;RL}, 
C_{\mu \mu }^{V\;RR}, C_{\mu \mu }^{V\;RL}, 
C_{\tau \tau }^{V\;RR}, C_{\tau \tau }^{V\;RL}, \nonumber \\
& 
C_{dd}^{V\;RR}, C_{dd}^{V\;RL}, 
C_{ss}^{V\;RR}, C_{ss}^{V\;RL}, 
C_{bb}^{V\;RR}, C_{bb}^{V\;RL}, 
C_{uu}^{V\;RR}, C_{uu}^{V\;RL}, 
C_{cc}^{V\;RR}, C_{cc}^{V\;RL} \big) \, ,
 \nonumber
\end{align}
with an analogous expression for the $L\leftrightarrow R$ interchange.
The structure of the solution to the RGE, \Eqn{RGE}, can then be written as
\begin{align}
\left(\begin{array}{c}
C^D_L(m_\mu) \\[8pt] \vec{C}_l^{ST}(m_\mu) \\[8pt]
\vec{C}_q^{ST}(\mu_n)\\[8pt] \vec{C}^V(m_\mu) \\[8pt] C_{gg}^L(\mu_n)
\end{array} \right)
  &=
\left(\begin{array}{c@{\extracolsep{8pt}}c@{\extracolsep{8pt}}c@{\extracolsep{8pt}}c@{\extracolsep{8pt}}c}
U_{DD} & U_{Dl}^{1\times 5} & 
U_{Dq}^{1\times 15}& U_{DV}^{1\times 16} &
U_{Dg} \\[8pt]
0^{5\times 1}  & U_{ll}^{5\times 5} & 0^{5\times 15} & 
0^{5\times 16} & 0^{5\times 1}\\[8pt]
0^{15\times 1}  & 0^{15\times 5}  & U_{qq}^{15\times 15} & 
0^{15\times 16} & U_{qg}^{15\times 1} \\[8pt]
0^{16\times 1}  & 0^{16\times 5} & 0^{16\times 15} &
U_{VV}^{16\times 16} & 0^{16\times 1} \\[8pt]
0 & 0^{1\times 5} & 0^{1\times 15} & 0^{1\times 16} & U_{gg}
\end{array} \right) \cdot
\left(\begin{array}{c}
C^D_L (m_W)\\[8pt] \vec{C}_l^{ST}(m_W) \\[8pt]
\vec{C}_q^{ST}(m_W)\\[8pt] \vec{C}^V(m_W) \\[8pt] C_{gg}^L(m_W) 
\end{array} \right)
\label{rge:num}
\end{align}
The choice of the low scale in the various parts of $\vec{C}$ on the
l.h.s. of \Eqn{rge:num} and the evolution from $\mu_n$ to $m_\mu$
deserves some comments\footnote{Here we remark that the different running scales are introduced in the present discussion for illustrative purpose only. In the phenomenological analysis of Section~\ref{sec:pheno} we have consistently extended the running of all Wilson coefficients to the physical scale of the process, namely $m_\mu$ for $\mu\to e\gamma$ and $\mu\to 3e$ and $\mu_n$ for $\mu\to e$ in nuclei.}}.

We start by noting that $C^D_L$, the first entry of $\vec{C}_l^{ST}$
and the first two entries of $\vec{C}^V$ contribute to either $\mu\to
e \gamma$ or $\mu\to3e$, \Eqns{muegBR}{br:meee}.  Hence, they have to
be evolved down to the scale $\mu=m_\mu$, relevant for these
processes. Since this scale is not in the perturbative domain of QCD
this is potentially problematic. Fortunately, none of the
corresponding anomalous dimensions receives contributions from
$\Gamma^s$. This leaves us with contributions from light-quark
operators which {\emph{per se}} are not meaningful at a scale below
$\mu_n=1\GeV$. However, the mixing of the corresponding tensor
operators into the dipole operator, \Eqn{rgedipole}, is suppressed by
light-quark masses. Furthermore, the mixing of light-quark vector
operators into $C_{ee}^V$, \Eqns{appVRR}{appVRL}, relevant for $\mu\to
3e$ modifies $C_{ee}^V(m_\mu)$ only marginally.  This leaves us with
the two-loop contributions of the light-quark vector operators to the
dipole. While the entries of the corresponding evolution matrix
$U_{DV}^{1\times16}$ does depend on how the evolution from $\mu_n$ to
$m_\mu$ is treated, these effects overall are so small that they do
not noticeably affect the limits we will present in
Section~\ref{sec:pheno}.  Hence we can safely stop the evolution of
the light-quark operators at $\mu_n=1$~GeV, even those that are part
of $\vec{C}^V(m_\mu)$.

Regarding the remaining two coefficients on the l.h.s. of
\Eqn{rge:num} we remark that only $\vec{C}_q^{ST}$ receives
contributions involving $\alpha_s$. Furthermore, $\vec{C}_q^{ST}$
contributes to $\mu\to e$ conversion, \Eqn{tildeCSL}. As discussed in
Section~\ref{sec:processes}, the relevant scale for this process is
$\mu_n$. A similar remark applies to $C_{gg}^L$ even though this
coefficient does not run at all. Thus, the evolution of
$\vec{C}_q^{ST}$ below the scale $\mu_n=1\GeV$ is neither possible,
nor required.

All entries of $U_{DV}^{1\times 16}$ are two-loop contributions. The
last entry of $U_{Dl}^{1\times 5}$ and several entries of
$U_{Dq}^{1\times 15}$ also receive leading two-loop contributions. The
other contributions to the evolution matrices are one loop.

For the presentation of the evolution matrices in the remainder of
this section we will set the light-quark masses to zero and stop the
evolution of the light-quark operators at $\mu_n$. While this will
affect some of the entries of \Eqn{rge:num}, these choices have no
notable influence on the limits presented in Section~\ref{sec:pheno}.
With these conventions, the numerical values for the entries of
\Eqn{rge:num} are the following:
\begin{align}
U_{DD}(m_\mu,m_W) &= 0.94 \nonumber \\[5pt]
U_{Dl}^{1\times 5}(m_\mu,m_W)&= (-0.0002,\, -0.04,\, 0.008,\, 3.2,\, -0.001 )\nonumber \\[5pt]
U_{Dq}^{1\times 15}(m_\mu,m_W)&=(0,\, 0,\, 0,\, 0,\, 0,\, 0,\, 
0.003,\, 4.6,\, -0.0006,\, 0,\, 0,\, 0,\, 
0.007,\, -2.7,\, -0.0008)\nonumber \\[5pt]
U_{DV}^{1\times 16}(m_\mu,m_W)&= 10^{-4} \ (-3.1,\, 1.4,\, -3.1,\, 3.5, \,
-0.9,\, 0.8,\, -0.4,\, 0.3,\, -0.4,\, 0.3,\, \nonumber \\
& \hspace*{41pt}
-0.3,\, 0.2, \, -1.2,\, 1.5,\, -1.2,\, 1.4)\nonumber \\[5pt]
U_{qg}^{15\times 1}(\mu_n,m_W)&= (0,0,0,0,0,0,
-1.1\cdot10^{-6},-4.6\cdot 10^{-10},-1.1\cdot 10^{-6},
0,0,0,\nonumber \\
& \hspace*{15pt}
-7.\cdot 10^{-7},5.6\cdot 10^{-10},
-7.\cdot 10^{-7})^T\nonumber \\[5pt]
U_{Dg}(\mu_n,m_W)&= - 10^{-9}\nonumber \\[5pt]
U_{gg}(\mu_n,m_W)&= 1 
\end{align}
The zero entries in $U_{Dq}^{1\times 15}$ and $U_{qg}^{15\times 1}$
are due to the neglect of light-quark masses.

The larger matrices are given in table form for better
readability. For $U_{ll}^{5\times 5}(m_\mu,m_W)$ we have
\begin{equation*} 
\begin{array}{c|ccccc}
U_{ll}^{5\times 5} &
C_{ee}^{S\;LL} & C_{\mu\mu}^{S\;LL} &
C_{\tau\tau}^{S\;LL} & C_{\tau\tau}^{T\;LL} & C_{\tau\tau}^{S\;RL} \\
\hline
C_{ee}^{S\;LL} & 0.95& 0.& 0.& 0.& 0. \\
C_{\mu\mu}^{S\;LL} & 0.& 0.95& 0.& 0.& 0. \\
C_{\tau\tau}^{S\;LL} & 0.& 0.& 1.03& 0.23& 0.  \\
C_{\tau\tau}^{T\;LL} & 0.& 0.& 0.005& 0.99& 0.  \\
C_{\tau\tau}^{S\;RL} &  0.& 0.& 0.& 0.& 1.03
\end{array}
\end{equation*}
whereas for $U_{qq}^{15\times 15}(\mu_n,m_W)$ the entries read
\begin{equation*} 
\resizebox{.99\linewidth}{!}{$\displaystyle
\begin{array}{c|ccccccccccccccc}
U_{qq}^{15\times 15}&
C_{dd}^{S\;LL}& C_{dd}^{T\;LL}& C_{dd}^{S\;RL}& 
C_{ss}^{S\;LL}& C_{ss}^{T\;LL}& C_{ss}^{S\;RL}& 
C_{bb}^{S\;LL}& C_{bb}^{T\;LL}& C_{bb}^{S\;RL}& 
C_{uu}^{S\;LL}& C_{uu}^{T\;LL}& C_{uu}^{S\;RL}& 
C_{cc}^{S\;LL}& C_{cc}^{T\;LL}& C_{cc}^{S\;RL}\\
\hline
C_{dd}^{S\;LL}& 1.85& 0.12& 0.& 0.& 0.& 0.& 0.& 0.& 0.& 0.& 0.& 0.& 0.& 0.& 0. \\
C_{dd}^{T\;LL}& 0.002& 0.82& 0.& 0.& 0.& 0.& 0.& 0.& 0.& 0.& 0.& 0.& 0.& 0.& 0. \\
C_{dd}^{S\;RL}& 0.& 0.& 1.85& 0.& 0.& 0.& 0.& 0.& 0.& 0.& 0.& 0.& 0.& 0.& 0. \\
C_{ss}^{S\;LL}& 0.& 0.& 0.& 1.85& 0.12& 0.& 0.& 0.& 0.& 0.& 0.& 0.& 0.& 0.& 0. \\
C_{ss}^{T\;LL}& 0.& 0.& 0.& 0.004& 0.65& 0.& 0.& 0.& 0.& 0.& 0.& 0.& 0.& 0.& 0. \\
C_{ss}^{S\;RL}& 0.& 0.& 0.& 0.& 0.& 1.85& 0.& 0.& 0.& 0.& 0.& 0.& 0.& 0.& 0. \\
C_{bb}^{S\;LL}& 0.& 0.& 0.& 0.& 0.& 0.& 1.38& 0.07& 0.& 0.& 0.& 0.& 0.& 0.& 0. \\
C_{bb}^{T\;LL}& 0.& 0.& 0.& 0.& 0.& 0.& 0.001& 0.90& 0.& 0.& 0.& 0.& 0.& 0.& 0. \\
C_{bb}^{S\;RL}& 0.& 0.& 0.& 0.& 0.& 0.& 0.& 0.& 1.38& 0.& 0.& 0.& 0.& 0.& 0. \\
C_{uu}^{S\;LL}& 0.& 0.& 0.& 0.& 0.& 0.& 0.& 0.& 0.& 1.86& -0.24& 0.& 0.& 0.& 0. \\
C_{uu}^{T\;LL}& 0.& 0.& 0.& 0.& 0.& 0.& 0.& 0.& 0.& -0.004& 0.81& 0.& 0.& 0.& 0. \\
C_{uu}^{S\;RL}& 0.& 0.& 0.& 0.& 0.& 0.& 0.& 0.& 0.& 0.& 0.& 1.86& 0.& 0.& 0. \\
C_{cc}^{S\;LL}& 0.& 0.& 0.& 0.& 0.& 0.& 0.& 0.& 0.& 0.& 0.& 0.& 1.81& -0.23& 0. \\
C_{cc}^{T\;LL}& 0.& 0.& 0.& 0.& 0.& 0.& 0.& 0.& 0.& 0.& 0.& 0.& -0.004& 0.82& 0. \\
C_{cc}^{S\;RL}& 0.& 0.& 0.& 0.& 0.& 0.& 0.& 0.& 0.& 0.& 0.& 0.& 0.& 0.& 1.81
\end{array}
$ 
}
\end{equation*}
Finally, the largest matrix is given in the form 
\begin{align}
U_{VV}^{16\times 16}(m_\mu,m_W) &= 
\mathbb{I} + 10^{-2}\, \bar{U}_{VV}(m_\mu,m_W)
\end{align}
and the entries of $\bar{U}_{VV}(m_\mu,m_W)$ are given by
\begin{equation*} 
\resizebox{.99\linewidth}{!}{$\displaystyle
\begin{array}{c|cccccccccccccccc}
\bar{U}_{VV} &
C_{ee}^{V\;RR} &
C_{ee}^{V\;RL}&
C_{\mu \mu }^{V\;RR}&  
C_{\mu \mu }^{V\;RL}& 
C_{\tau \tau }^{V\;RR}& 
C_{\tau \tau }^{V\;RL}&
C_{dd}^{V\;RR}&
C_{dd}^{V\;RL}& 
C_{ss}^{V\;RR}& 
C_{ss}^{V\;RL}& 
C_{bb}^{V\;RR}& 
C_{bb}^{V\;RL}& 
C_{uu}^{V\;RR}& 
C_{uu}^{V\;RL}& 
C_{cc}^{V\;RR}& 
C_{cc}^{V\;RL}\\
\hline
C_{ee}^{V\;RR} & -5.8& -0.5& -1.& -0.5& -0.3& -0.3& -0.3& -0.3& 
-0.3& -0.3& -0.2& -0.2& 0.7& 0.7& 0.7& 0.7 \\
C_{ee}^{V\;RL}&  -1.1& 4.4& -1.1& -0.6& -0.3& -0.3& -0.4& -0.4& -0.4& 
-0.4& -0.2& -0.3& 0.7& 0.7& 0.7& 0.7  \\
C_{\mu \mu }^{V\;RR}&  -1.& -0.5& -5.8& -0.5& -0.3& -0.3& -0.3& -0.3& -0.3& 
-0.3& -0.2& -0.2& 0.7& 0.7& 0.7& 0.7  \\
C_{\mu \mu }^{V\;RL}&  -1.1& -0.6& -1.1& 4.4& -0.3& -0.3& -0.4& -0.4& -0.4& 
-0.4& -0.2& -0.3& 0.7& 0.7& 0.7& 0.7  \\
C_{\tau \tau }^{V\;RR}& -0.6& -0.3& -0.6& -0.3& -3.1& -0.3& -0.3& -0.3& 
-0.3& -0.3& -0.2& -0.2& 0.6& 0.6& 0.6& 0.6   \\
C_{\tau \tau }^{V\;RL}& -0.6& -0.3& -0.6& -0.3& -0.3& 2.6& -0.3& -0.3& 
-0.3& -0.3& -0.2& -0.2& 0.6& 0.6& 0.6& 0.6   \\
C_{dd}^{V\;RR}& -0.2& -0.1& -0.2& -0.1& -0.1& -0.1& -1.2& -0.1& 
-0.1& -0.1& -0.1& -0.1& 0.2& 0.2& 0.2& 0.2   \\
C_{dd}^{V\;RL}& -0.2& -0.1& -0.2& -0.1& -0.1& -0.1& -0.1& 1.0& 
-0.1& -0.1& -0.1& -0.1& 0.2& 0.2& 0.2& 0.2   \\
C_{ss}^{V\;RR}& -0.2& -0.1& -0.2& -0.1& -0.1& -0.1& -0.1& -0.1& 
-1.2& -0.1& -0.1& -0.1& 0.2& 0.2& 0.2& 0.2   \\
C_{ss}^{V\;RL}&  -0.2& -0.1& -0.2& -0.1& -0.1& -0.1& -0.1& -0.1& 
-0.1& 1.0& -0.1& -0.1& 0.2& 0.2& 0.2& 0.2  \\
C_{bb}^{V\;RR}& -0.2& -0.1& -0.2& -0.1& -0.1& -0.1& -0.1& -0.1& 
-0.1& -0.1& -0.8& -0.1& 0.2& 0.2& 0.2& 0.2   \\
C_{bb}^{V\;RL}&  -0.2& -0.1& -0.2& -0.1& -0.1& -0.1& -0.1& -0.1& 
-0.1& -0.1& -0.1& 0.7& 0.2& 0.2& 0.2& 0.2  \\
C_{uu}^{V\;RR}& 0.5& 0.2& 0.5& 0.2& 0.2& 0.2& 0.2& 0.2& 0.2& 0.2& 
0.2& 0.2& 1.7& -0.5& -0.5& -0.5   \\
C_{uu}^{V\;RL}& 0.5& 0.2& 0.5& 0.2& 0.2& 0.2& 0.2& 0.2& 0.2& 0.2& 
0.2& 0.2& -0.5& -2.6& -0.5& -0.5   \\
C_{cc}^{V\;RR}&  0.5& 0.2& 0.5& 0.2& 0.2& 0.2& 0.2& 0.2& 0.2& 0.2& 
0.2& 0.2& -0.5& -0.5& 1.7& -0.5  \\
C_{cc}^{V\;RL}& 0.5& 0.2& 0.5& 0.2& 0.2& 0.2& 0.2& 0.2& 0.2& 0.2& 
0.2& 0.2& -0.5& -0.5& -0.5& -2.6
\end{array}
$ 
}
\end{equation*}  
Of course, it is no problem to change the high scale from $m_W$ to
another scale $\Lambda$, as long as $m_W\ge \Lambda\gg m_b$.


\section{Phenomenological analysis \label{sec:pheno}}

In this section we use the concepts presented previously for a
phenomenological analysis\footnote{Parts of these results were already
  presented in the proceedings~\cite{Crivellin:2016ebg}}. For this
purpose, we assume that the Wilson coefficients (which are generated
by some underlying NP theory {above the EWSB scale}) are given at {$\Lambda=m_W$}. In a
first step, we use the RGEs presented in Section~\ref{sec:running}
to evolve the Wilson coefficients from the high scale $m_W$ to the low
experimental scale (namely $\mu_n$ and $m_\mu$, respectively). At the
low scale, the predicted rates are then confronted with the
experimental limits. This results in constraints on the various Wilson
coefficients at the high scale. Due to mixing effects in the RGE,
Wilson coefficients that are zero at the high scale can be non-zero at
the low scale. Hence we are able to place bounds on coefficients which
would be unconstrained if loop effects were disregarded.

In this analysis, we use the following current experimental limits:
\begin{eqnarray}
	&&{\rm Br}\left( \mu^+ \to e^+ \gamma \right)\leq4.2 \times10^{-13}\,,\\
	&&{\rm Br}\left(\mu^+\to e^+e^-e^+ \right)\leq1.0 \times10^{-12}\,,\\
	&&{\rm Br}_{\mu \to e}^{\rm Au} \leq 7.0 \times 10^{-13}\,,
\end{eqnarray}
from the MEG~\cite{TheMEG:2016wtm,Adam:2013mnn},
SINDRUM~\cite{Bellgardt:1987du} and SINDRUM~II~\cite{Bertl:2006up}
collaborations, respectively. In addition, the following future
experimental limits
\begin{eqnarray}
	&&{\rm Br}\left( \mu^+ \to e^+ \gamma \right)\leq4.0 \times10^{-14}\,,\\
	&&{\rm Br}\left(\mu^+\to e^+e^-e^+ \right)\leq 5.0 \times10^{-15}\,,\\
	&&{\rm Br}_{\mu \to e}^{\rm Al} \leq 1.0 \times 10^{-16}\,,
\end{eqnarray}
from the MEG~II~\cite{Baldini:2013ke}, Mu3e~\cite{Blondel:2013ia} and
Mu2e/COMET~\cite{Carey:2008zz,Kutschke:2011ux,Cui:2009zz} are
considered.

\begin{table}[!t]
\begin{center}
\begin{tabular}{c||cc|cc|cc}
  & \multicolumn{2}{c|}{${\rm Br}\left( \mu^+ \to e^+ \gamma \right)$}
  & \multicolumn{2}{c|}{${\rm Br}\left(\mu^+\to e^+e^-e^+\right)$}
  & \multicolumn{2}{c}{${\rm Br}_{\mu \to e}^{\rm Au / Al}$ } \\
  & $4.2\cdot 10^{-13}$ & $4.0\cdot 10^{-14}$ &  $1.0\cdot 10^{-12}$
  & $5.0\cdot 10^{-15}$ &  $7.0\cdot 10^{-13}$ & $1.0\cdot 10^{-16}$
  \\[3pt]
 \hline\\[-10pt]
 $C^D_L$ & $1.0\cdot 10^{-8}$ & $3.1\cdot 10^{-9}$ & $2.0\cdot 10^{-7}$ & $1.4\cdot 10^{-8}$ & $2.0\cdot 10^{-7}$ & $2.9\cdot 10^{-9}$ \\
 $C_{ee}^{S\; LL}$ & $4.8\cdot 10^{-5}$ & $1.5\cdot 10^{-5}$ & $8.1\cdot 10^{-7}$ & $5.8\cdot 10^{-8}$ & $1.4\cdot 10^{-3}$ & $2.1\cdot 10^{-5}$ \\
 $C_{\mu\mu}^{S\; LL}$ & $2.3\cdot 10^{-7}$ & $7.2\cdot 10^{-8}$ & $4.6\cdot 10^{-6}$ & $3.3\cdot 10^{-7}$ & $7.1\cdot 10^{-6}$ & $1.0\cdot 10^{-7}$ \\
 $C_{\tau\tau}^{S\; LL}$ & $1.2\cdot 10^{-6}$ & $3.7\cdot 10^{-7}$ & $2.4\cdot 10^{-5}$ & $1.7\cdot 10^{-6}$ & $2.4\cdot 10^{-5}$ & $3.5\cdot 10^{-7}$ \\
 $C_{\tau\tau}^{T\; LL}$ & $2.9\cdot 10^{-9}$ & $9.0\cdot 10^{-10}$ & $5.7\cdot 10^{-8}$ & $4.1\cdot 10^{-9}$ & $5.9\cdot 10^{-8}$ & $8.5\cdot 10^{-10}$ \\
 $C_{\tau\tau}^{S\; LR}$ & $9.4\cdot 10^{-6}$ & $2.9\cdot 10^{-6}$ & $1.8\cdot 10^{-4}$ & $1.3\cdot 10^{-5}$ & $1.9\cdot 10^{-4}$ & $2.7\cdot 10^{-6}$ \\
 $C_{bb}^{S\; LL}$ & $2.8\cdot 10^{-6}$ & $8.6\cdot 10^{-7}$ & $5.4\cdot 10^{-5}$ & $3.8\cdot 10^{-6}$ & $9.0\cdot 10^{-7}$ & $1.2\cdot 10^{-8}$ \\
 $C_{bb}^{T\; LL}$ & $2.1\cdot 10^{-9}$ & $6.4\cdot 10^{-10}$ & $4.1\cdot 10^{-8}$ & $2.9\cdot 10^{-9}$ & $4.2\cdot 10^{-8}$ & $6.0\cdot 10^{-10}$ \\
 $C_{bb}^{S\; LR}$ & $1.7\cdot 10^{-5}$ & $5.1\cdot 10^{-6}$ & $3.2\cdot 10^{-4}$ & $2.3\cdot 10^{-5}$ & $9.1\cdot 10^{-7}$ & $1.2\cdot 10^{-8}$ \\
 $C_{cc}^{S\; LL}$ & $1.4\cdot 10^{-6}$ & $4.4\cdot 10^{-7}$ & $2.8\cdot 10^{-5}$ & $2.0\cdot 10^{-6}$ & $1.8\cdot 10^{-7}$ & $2.4\cdot 10^{-9}$ \\
 $C_{cc}^{T\; LL}$ & $3.5\cdot 10^{-9}$ & $1.1\cdot 10^{-9}$ & $6.8\cdot 10^{-8}$ & $4.8\cdot 10^{-9}$ & $6.6\cdot 10^{-8}$ & $9.5\cdot 10^{-10}$ \\
 $C_{cc}^{S\; LR}$ & $1.2\cdot 10^{-5}$ & $3.6\cdot 10^{-6}$ & $2.3\cdot 10^{-4}$ & $1.6\cdot 10^{-5}$ & $1.8\cdot 10^{-7}$ & $2.4\cdot 10^{-9}$ \\
 $C_{ee}^{V\; RR}$ & $3.0\cdot 10^{-5}$ & $9.4\cdot 10^{-6}$ & $2.1\cdot 10^{-7}$ & $1.5\cdot 10^{-8}$ & $2.1\cdot 10^{-6}$ & $3.5\cdot 10^{-8}$ \\
 $C_{ee}^{V\; RL}$ & $6.7\cdot 10^{-5}$ & $2.1\cdot 10^{-5}$ & $2.6\cdot 10^{-7}$ & $1.9\cdot 10^{-8}$ & $4.0\cdot 10^{-6}$ & $6.7\cdot 10^{-8}$ \\
 $C_{\mu\mu}^{V\; RR}$ & $3.0\cdot 10^{-5}$ & $9.4\cdot 10^{-6}$ & $1.6\cdot 10^{-5}$ & $1.1\cdot 10^{-6}$ & $2.1\cdot 10^{-6}$ & $3.5\cdot 10^{-8}$ \\
 $C_{\mu\mu}^{V\; RL}$ & $2.7\cdot 10^{-5}$ & $8.5\cdot 10^{-6}$ & $2.9\cdot 10^{-5}$ & $2.0\cdot 10^{-6}$ & $4.0\cdot 10^{-6}$ & $6.6\cdot 10^{-8}$ \\
 $C_{\tau\tau}^{V\; RR}$ & $1.0\cdot 10^{-4}$ & $3.2\cdot 10^{-5}$ & $5.3\cdot 10^{-5}$ & $3.8\cdot 10^{-6}$ & $4.8\cdot 10^{-6}$ & $7.9\cdot 10^{-8}$ \\
 $C_{\tau\tau}^{V\; RL}$ & $1.2\cdot 10^{-4}$ & $3.6\cdot 10^{-5}$ & $5.1\cdot 10^{-5}$ & $3.6\cdot 10^{-6}$ & $4.6\cdot 10^{-6}$ & $7.6\cdot 10^{-8}$ \\
 $C_{bb}^{V\; RR}$ & $3.5\cdot 10^{-4}$ & $1.1\cdot 10^{-4}$ & $6.7\cdot 10^{-5}$ & $4.8\cdot 10^{-6}$ & $6.0\cdot 10^{-6}$ & $1.0\cdot 10^{-7}$ \\
 $C_{bb}^{V\; RL}$ & $5.3\cdot 10^{-4}$ & $1.6\cdot 10^{-4}$ & $6.6\cdot 10^{-5}$ & $4.7\cdot 10^{-6}$ & $6.0\cdot 10^{-6}$ & $9.9\cdot 10^{-8}$ \\
 $C_{cc}^{V\; RR}$ & $8.1\cdot 10^{-5}$ & $2.5\cdot 10^{-5}$ & $2.3\cdot 10^{-5}$ & $1.6\cdot 10^{-6}$ & $2.1\cdot 10^{-6}$ & $3.4\cdot 10^{-8}$ \\
 $C_{cc}^{V\; RL}$ & $6.7\cdot 10^{-5}$ & $2.1\cdot 10^{-5}$ & $2.4\cdot 10^{-5}$ & $1.7\cdot 10^{-6}$ & $2.1\cdot 10^{-6}$ & $3.5\cdot 10^{-8}$ \\
 $C_{gg}^L$ & N/A & N/A & N/A & N/A & $6.2\cdot 10^{-3}$ & $8.1\cdot 10^{-5}$ \\
\end{tabular}
\end{center}
\caption{\label{tab:cc} Limits on the various coefficients $C_i(m_W)$
  from current and future experimental constraints, assuming that (at
  the high scale $m_W$) only one coefficient at a time is
  non-vanishing and not including operator-dependent efficiency
  corrections.}
\end{table}

Let us start by assuming that at the high scale $m_W$ only one Wilson
coefficient at a time is non-zero. The corresponding bounds on the
coefficients are shown in Table~\ref{tab:cc} both for the current and
for the future experimental limits. Note that for $\mu\to 3e$ we did
not take into account efficiency corrections due to cuts on the
transverse momentum applied in the experimental analysis. These
corrections are in general small for 4-fermion operators but
significantly reduce the sensitivity to the dipole operator. From
Table~\ref{tab:cc} we can infer the following general structure of
these limits:
\begin{itemize}
\item Experimental bounds on the direct $\mu\to e\gamma$ transition
  represent a powerful tool to test the Wilson coefficients of the
  dipole operator. Furthermore, the impact of mixing effects
  originating from some scalar and tensor operators can also be
  examined with high precision. However, future prospects for nuclear
  conversion are so good, that it could overtake the direct $\mu\to
  e\gamma$ limits. The only (numerically accidental) exception is
  represented by $C^{S}_{\mu\mu}$, that will be still better
  constrained by the next generation of $\mu\to e\gamma$ experiments.
\item A $\mu\to 3e$ experiment is the most powerful tool to explore
  $\mu$-$e$-$e$-$e$ Wilson coefficients of four fermion operators,
  regardless of the Dirac structure of the operator. This is mainly
  due to the fact that such interactions produce the $\mu\to 3e$ decay
  already at the tree level (see Eq.~(\ref{br:meee})) while it enters
  all other processes only via loop effects.
\item As expected, $\mu\to e$ conversion is the most sensitive
  experimental framework to explore the set of operators including
  quarks (with exception of current limits on $C^{T}_{cc/bb}$, the
  Wilson coefficients of tensor operators) and gluons. However, it
  also appears to be the best setup to study any kind of vector
  interaction (with the exception of the aforementioned $C^{V}_{ee}$
  operators, for which $\mu\to 3e$ represents the golden
  channel). This is mostly due to notable RGE effects in the vector
  operator mixing matrix.
\end{itemize}

\begin{table}[!t]
\begin{center}
\begin{tabular}{c||cc|cc|cc}
  & \multicolumn{2}{c|}{${\rm Br}\left( \mu^+ \to e^+ \gamma \right)$}
  & \multicolumn{2}{c|}{${\rm Br}\left(\mu^+\to e^+e^-e^+\right)$}
  & \multicolumn{2}{c}{${\rm Br}_{\mu \to e}^{\rm Au / Al}$ } \\
  & $4.2\cdot 10^{-13}$ & $4.0\cdot 10^{-14}$ &  $1.0\cdot 10^{-12}$
  & $5.0\cdot 10^{-15}$ &  $7.0\cdot 10^{-13}$ & $1.0\cdot 10^{-16}$
  \\[3pt]
 \hline\\[-10pt]
 $C^{LP}_{bb}$ & $4.7\cdot 10^{-6}$ & $1.5\cdot 10^{-6}$ & $9.3\cdot 10^{-5}$ & $6.6\cdot 10^{-6}$ & $9.6\cdot 10^{-5}$ & $1.4\cdot 10^{-6}$ \\
 $C^{LS}_{bb}$ & $6.7\cdot 10^{-6}$ & $2.1\cdot 10^{-6}$ & $1.3\cdot 10^{-4}$ & $9.2\cdot 10^{-6}$ & $9.1\cdot 10^{-7}$ & $1.2\cdot 10^{-8}$ \\
 $C^{LP}_{cc}$ & $2.6\cdot 10^{-6}$ & $7.9\cdot 10^{-7}$ & $5.0\cdot 10^{-5}$ & $3.5\cdot 10^{-6}$ & $5.0\cdot 10^{-5}$ & $7.1\cdot 10^{-7}$ \\
 $C^{LS}_{cc}$ & $3.3\cdot 10^{-6}$ & $1.0\cdot 10^{-6}$ & $6.4\cdot 10^{-5}$ & $4.5\cdot 10^{-6}$ & $1.8\cdot 10^{-7}$ & $2.4\cdot 10^{-9}$ \\
{ $C^{RA}_{dd}$} & $2.9\cdot 10^{-4}$ & $8.8\cdot 10^{-5}$ & $3.9\cdot 10^{-3}$ & $2.8\cdot 10^{-4}$ & $6.6\cdot 10^{-7}$ & $1.4\cdot 10^{-8}$ \\
{ $C^{RV}_{dd}$} & $1.4\cdot 10^{-3}$ & $4.3\cdot 10^{-4}$ & $4.5\cdot 10^{-5}$ & $3.2\cdot 10^{-6}$ & $7.2\cdot 10^{-9}$ & $1.5\cdot 10^{-10}$ \\ì
{ $C^{RA}_{ss}$} & $2.9\cdot 10^{-4}$ & $8.8\cdot 10^{-5}$ & $3.9\cdot 10^{-3}$ & $2.8\cdot 10^{-4}$ & $6.5\cdot 10^{-4}$ & $1.1\cdot 10^{-5}$ \\
{ $C^{RV}_{ss}$} & $1.4\cdot 10^{-3}$ & $4.3\cdot 10^{-4}$ & $4.5\cdot 10^{-5}$ & $3.2\cdot 10^{-6}$ & $4.1\cdot 10^{-6}$ & $6.8\cdot 10^{-8}$ \\
 $C^{RA}_{bb}$ & $4.2\cdot 10^{-4}$ & $1.3\cdot 10^{-4}$ & $6.5\cdot 10^{-3}$ & $4.6\cdot 10^{-4}$ & $1.3\cdot 10^{-3}$ & $2.2\cdot 10^{-5}$ \\
 $C^{RV}_{bb}$ & $2.1\cdot 10^{-3}$ & $6.4\cdot 10^{-4}$ & $6.7\cdot 10^{-5}$ & $4.7\cdot 10^{-6}$ & $6.0\cdot 10^{-6}$ & $1.0\cdot 10^{-7}$ \\
{ $C^{RA}_{uu}$} & $7.2\cdot 10^{-5}$ & $2.2\cdot 10^{-5}$ & $9.8\cdot 10^{-4}$ & $6.9\cdot 10^{-5}$ & $3.8\cdot 10^{-7}$ & $7.2\cdot 10^{-9}$ \\
{ $C^{RV}_{uu}$} & $7.7\cdot 10^{-4}$ & $2.4\cdot 10^{-4}$ & $2.3\cdot 10^{-5}$ & $1.6\cdot 10^{-6}$ & $8.3\cdot 10^{-9}$ & $1.6\cdot 10^{-10}$ \\
 $C^{RA}_{cc}$ & $7.3\cdot 10^{-5}$ & $2.3\cdot 10^{-5}$ & $1.0\cdot 10^{-3}$ & $7.1\cdot 10^{-5}$ & $1.7\cdot 10^{-4}$ & $2.8\cdot 10^{-6}$ \\
 $C^{RV}_{cc}$ & $7.8\cdot 10^{-4}$ & $2.4\cdot 10^{-4}$ & $2.3\cdot 10^{-5}$ & $1.6\cdot 10^{-6}$ & $2.1\cdot 10^{-6}$ & $3.5\cdot 10^{-8}$
\end{tabular}
\end{center}
\caption{\label{tab:VA} Limits on the Wilson coefficients (given at
  the scale $m_W$) with scalar (vector) and pseudo-scalar
  (axial-vector) currents from current and future experimental
  constraints, assuming that only one coefficient at a time is
  non-vanishing at the high scale.}
\end{table}

\begin{figure}[!th]
\begin{center}
\subfloat[]{
\includegraphics[width=0.24\textwidth]{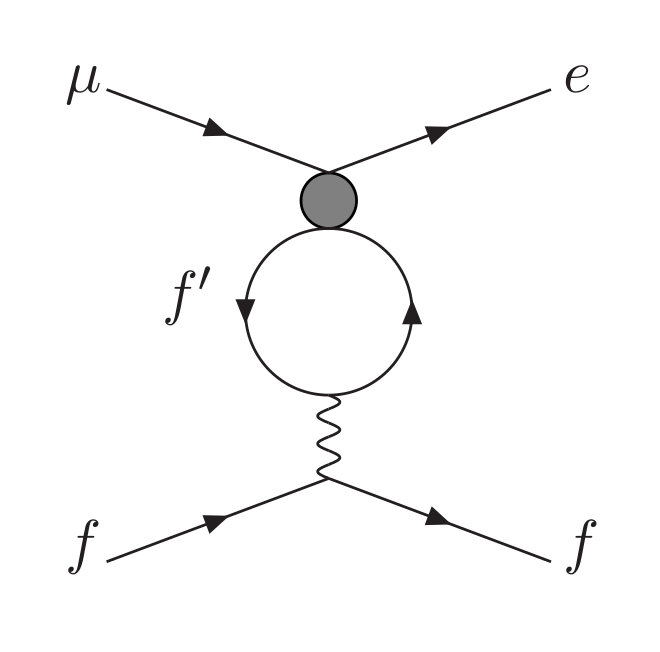}
\label{fig:feyna}
}
\subfloat[]{
\includegraphics[width=0.24\textwidth]{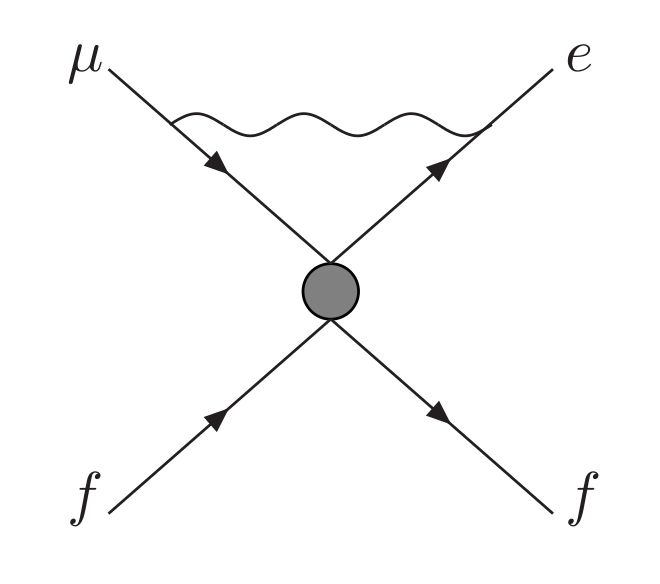}
\includegraphics[width=0.24\textwidth]{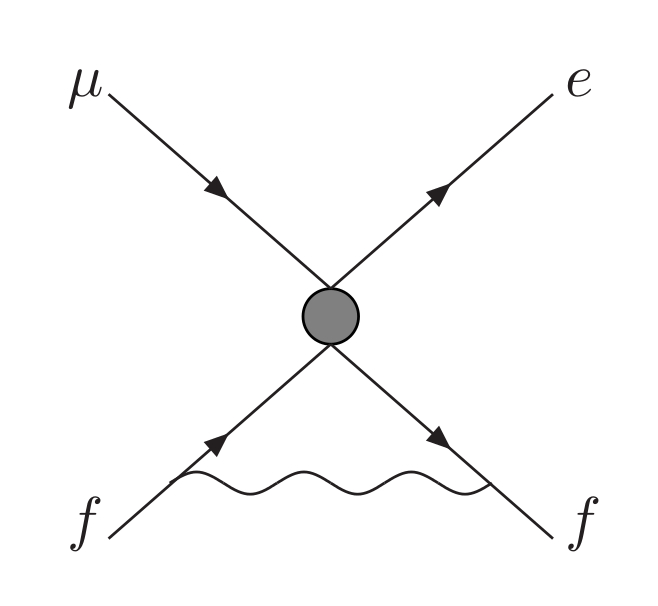}
\label{fig:feynb}
}\\
\subfloat[]{
\includegraphics[width=0.24\textwidth]{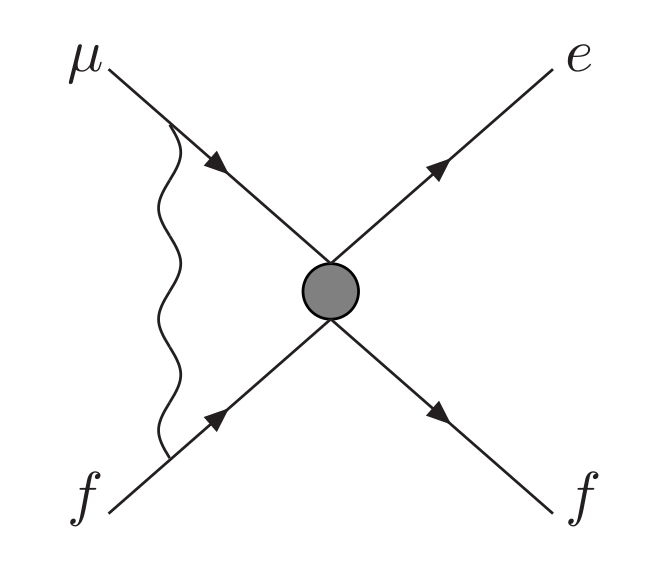}
\includegraphics[width=0.24\textwidth]{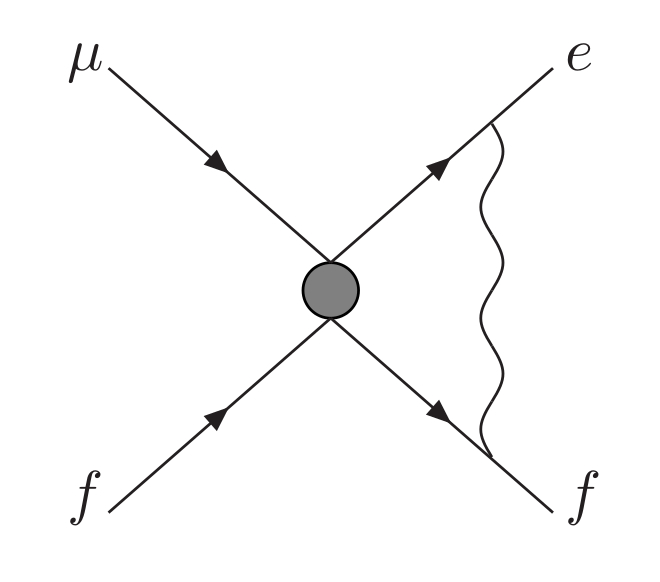}
\includegraphics[width=0.24\textwidth]{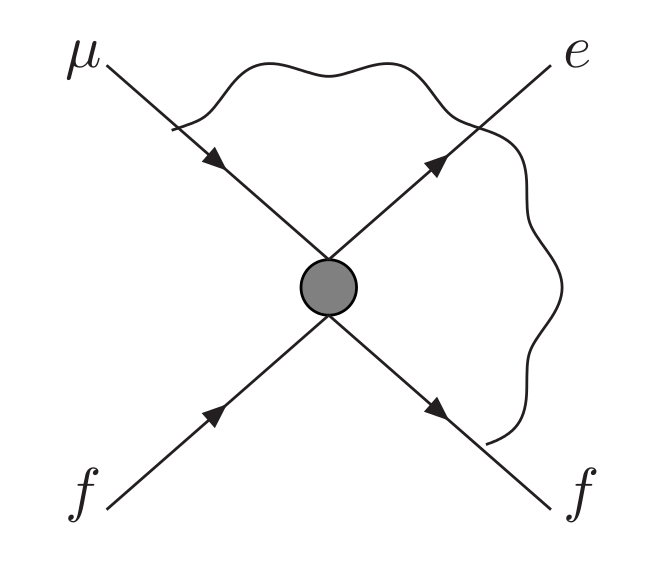}
\includegraphics[width=0.24\textwidth]{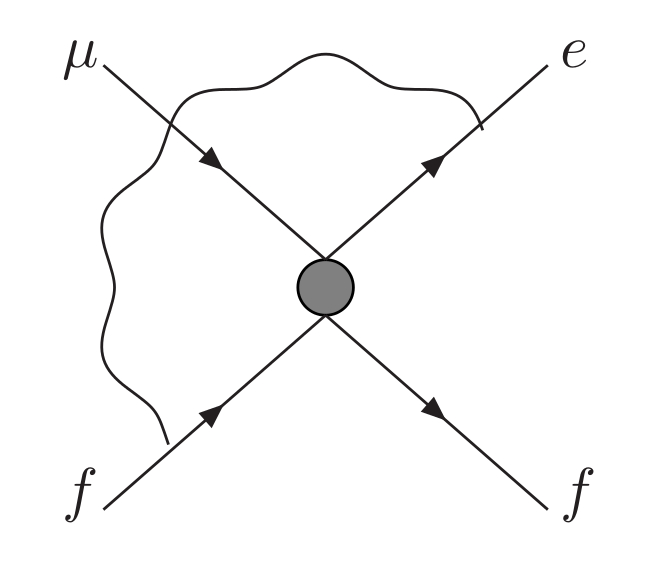}
\label{fig:feync}
}
\end{center}
\caption{One-loop Feynman diagrams for the mixing of the four-fermion lepton-quark and lepton-tau operators.}
\label{fig:feyn}
\end{figure}

Concerning, $\mu\to e$ conversion it is important to keep in mind that
we chose for the constraints in Table~\ref{tab:cc} a chiral basis,
\emph{i.e.} we worked with left- and right-handed fields. However, for
Wilson coefficients given at the low experimental scale, the $\mu\to
e$ conversion rate is only sensitive to operators with vector or
scalar currents on the quark side, but not to operators with
axial-vector or pseudo-scalar currents. Therefore, it is informative
to switch the basis and consider operators with scalar (vector) and
pseudo-scalar (axial-vector) currents instead:
\begin{align}
\hspace*{2cm}
C^{XS}_{ff}&=\frac{C^{S XR}_{ff}+C^{S XL}_{ff}}{2}, & 
C^{XV}_{ff}&=\frac{C^{V XR}_{ff}+C^{V XL}_{ff}}{2}, \hspace*{2cm} \\
C^{XP}_{ff}&=\frac{C^{S XR}_{ff}-C^{S XL}_{ff}}{2}, &
C^{XA}_{ff}&=\frac{C^{V XR}_{ff}-C^{V XL}_{ff}}{2}, &
\end{align}
where $X\in\{L,R\}$ and $f\in\{u,c,d,s,b,e,\mu,\tau\}$. In a
simplistic tree-level approach, ${\mu\to e}$ conversion is not sensitive
to $C^{XA}_{ff}$ and $C^{XP}_{ff}$. However, the remarkable outcome is
that axial-vector operators mix into vector operators. This results
in strong bounds from ${\mu\to e}$ conversion in nuclei once the
Wilson coefficients are evaluated at a scale higher than the
experimental scale. Therefore, the common preconceptions that $\mu\to
e$ conversion is not sensitive to axial-vector currents is not true
anymore once loop effects are taken into account. The corresponding
results are presented in Table~\ref{tab:VA}. {In order to understand how the parity-selection rules work for the vectorial lepton-quark and lepton-tau operators, in Figure~\ref{fig:feyn} we show the Feynman diagrams of their mixing at the one-loop level.}

{First, vectorial operators do not receive contribution from the diagrams in Figure~\ref{fig:feynb} (they vanish when the wave-function renormalisation is included). After, we remark that the penguin diagram in Figure~\ref{fig:feyna} generates operators in which the flavour conserving current has a vectorial structure, \emph{i.e.} vector-vector and axial-vector operators. Then, from the contribution of the diagrams in Figure~\ref{fig:feync} one obtains a ``maximally flipped'' operatorial mixing: axial-axial into vector-vector, axial-vector into vector-axial, vector-axial into axial-vector and vector-vector into axial-axial. By combining the contributions, the parity selection rules work as follows: vector-vector and axial-axial operators mix into vector-vector (with the contribution from the penguin diagram), axial-vector mixes into vector-axial, axial-vector and vector-axial mix into axial-vector (with the contribution from the penguin diagram), and vector-vector mixes into axial-axial. These results have been discussed also by previous literature~\cite{Crivellin:2014qxa,DEramo:2016gos}.}

\begin{figure}[!th]
\begin{center}
\includegraphics[width=0.85\textwidth]{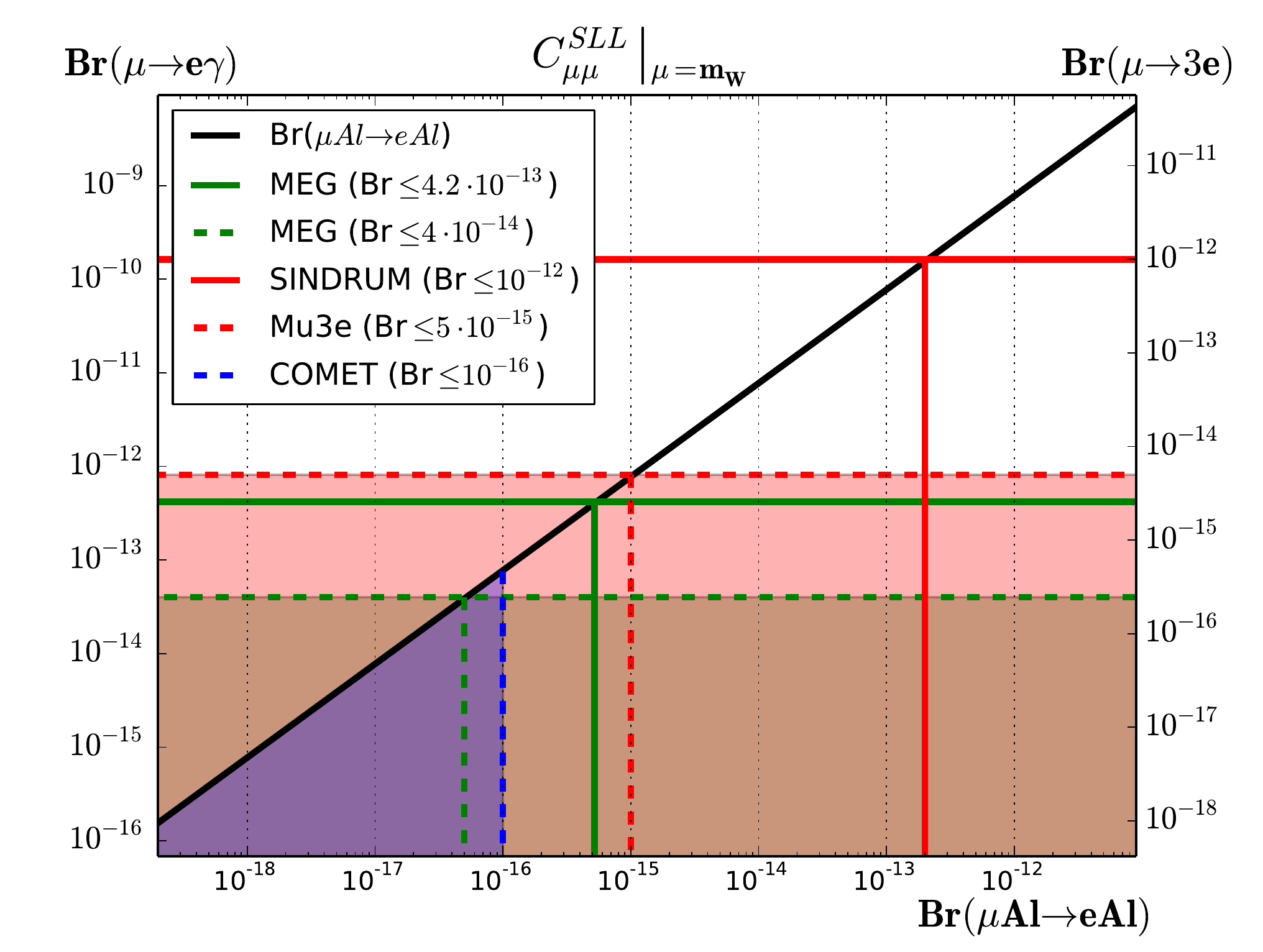}\\
\includegraphics[width=0.85\textwidth]{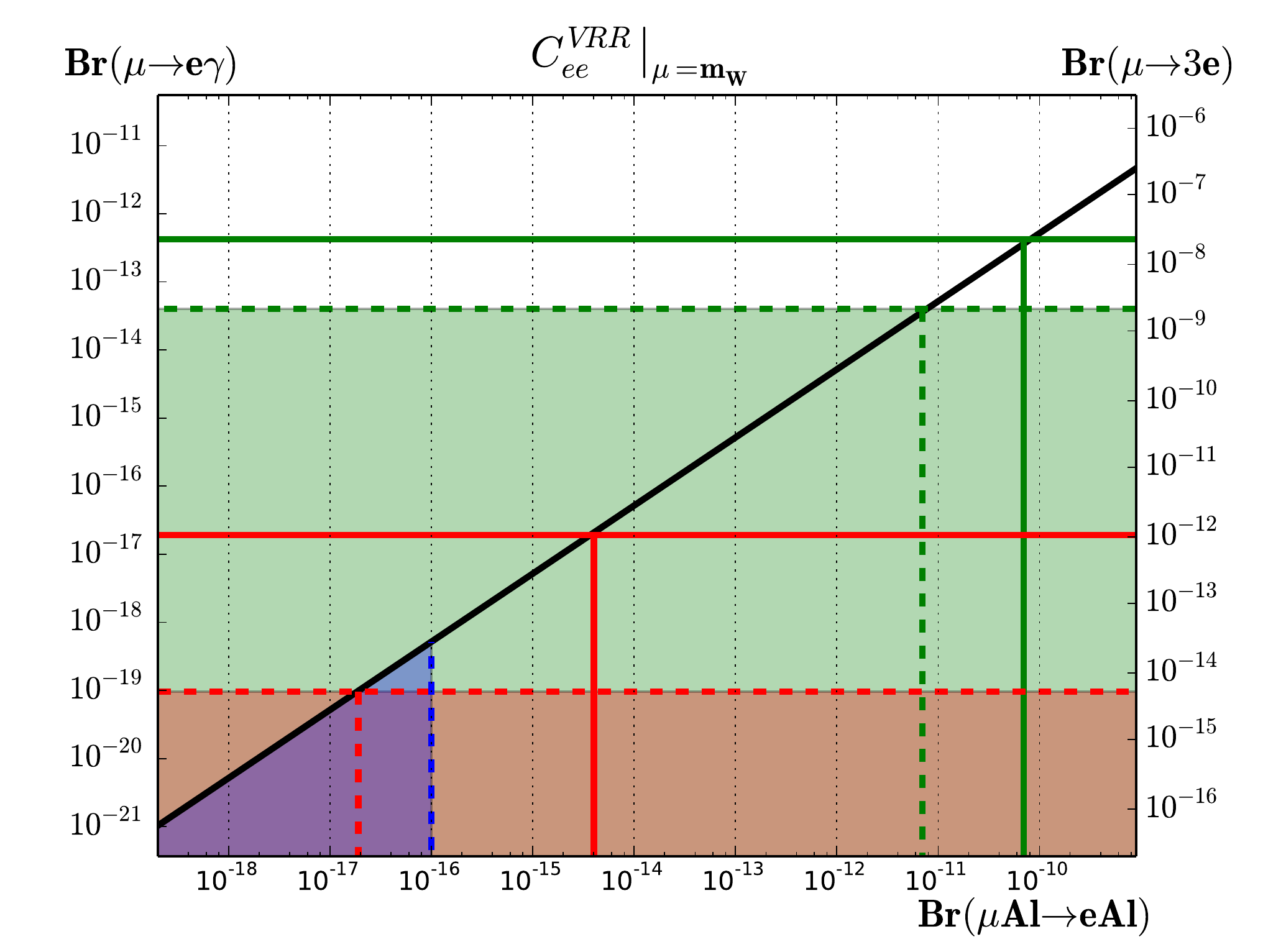}
\end{center}
\caption{\Br($\mu\to e\gamma$) (\Br($\mu\to 3e$)) plotted on the left
  (right) $y$-axis against \Br($\mu N\to eN$) for a fixed value of
  $C^{S\; LL}_{\mu\mu}$ (upper panel) and $C^{V\; RR}_{ee}$ (lower
  panel) given at the scale $\mu=m_W$. Current and future
  experimental limits are displayed.}
\label{figE}
\end{figure}

In a next step let us compare the exploring power of current and
future $\mu\to e\gamma$, $\mu\to 3e$ and $\mu\to e$ conversion
experiments by directly relating the branching-ratio limits that are
needed for the various processes to achieve a particular bound on a
Wilson coefficient. For illustrative purposes, we single out two
coefficients namely $C^{S\; LL}_{\mu\mu}$ and $C^{V\; RR}_{ee}$. In
Figure~\ref{figE}, the current and future branching ratio for $\mu\to
e\gamma$ and $\mu\to 3e$ experiments are compared to the future $\mu
N\to e N$ prospects (where $N$ is an aluminium nucleus).

Starting with the upper panel, the horizontal dashed red line for
example indicates that a rather modest limit $\Br(\mu\to
e\gamma)\lesssim 10^{-12}$ is as constraining on $C^{S\,LL}_{\mu\mu}$
as the future Mu3e limit $\Br(\mu\to 3e)< 5\times 10^{-15}$. In order
for muon conversion to be more constraining a limit of $\Br(\mu N\to
eN)< 10^{-15}$ would be required. This is indicated by the vertical
dashed red line. The future MEG~II experiment will place the strongest
limit on $C^{S\,LL}_{\mu\mu}$ unless the COMET or Mu2e experiments
improve their expected limit to reach at least $\Br(\mu N\to eN)<
5\times10^{-17}$. For this specific operator Mu3e will have less of an
impact.

From the lower panel, we infer that the future Mu3e experiment will
deliver the best bound on $C^{V\,RR}_{ee}$. In order to perform with
similar standards, the COMET and Mu2e experiments will have to obtain
a limit of $\Br(\mu N\to eN)< 2\times10^{-17}$. Instead,
in this specific context, the MEG~II experiment will exhibit a
generally weak sensitivity, orders of magnitude below the capability
of the other experiments.

So far we have considered a scenario where only one coefficient at a
time is non-vanishing at the high scale.  Now, let us extend the
previous results by assuming two Wilson coefficients are
non-vanishing. For this purpose, we generated plots in which the
parameter space is analysed in light of current and future
experimental limits for all three processes. For a better
understanding, we display them in a (pseudo-)logarithmic scale.

\begin{figure}[!th]
\begin{center}
\includegraphics[width=\textwidth]{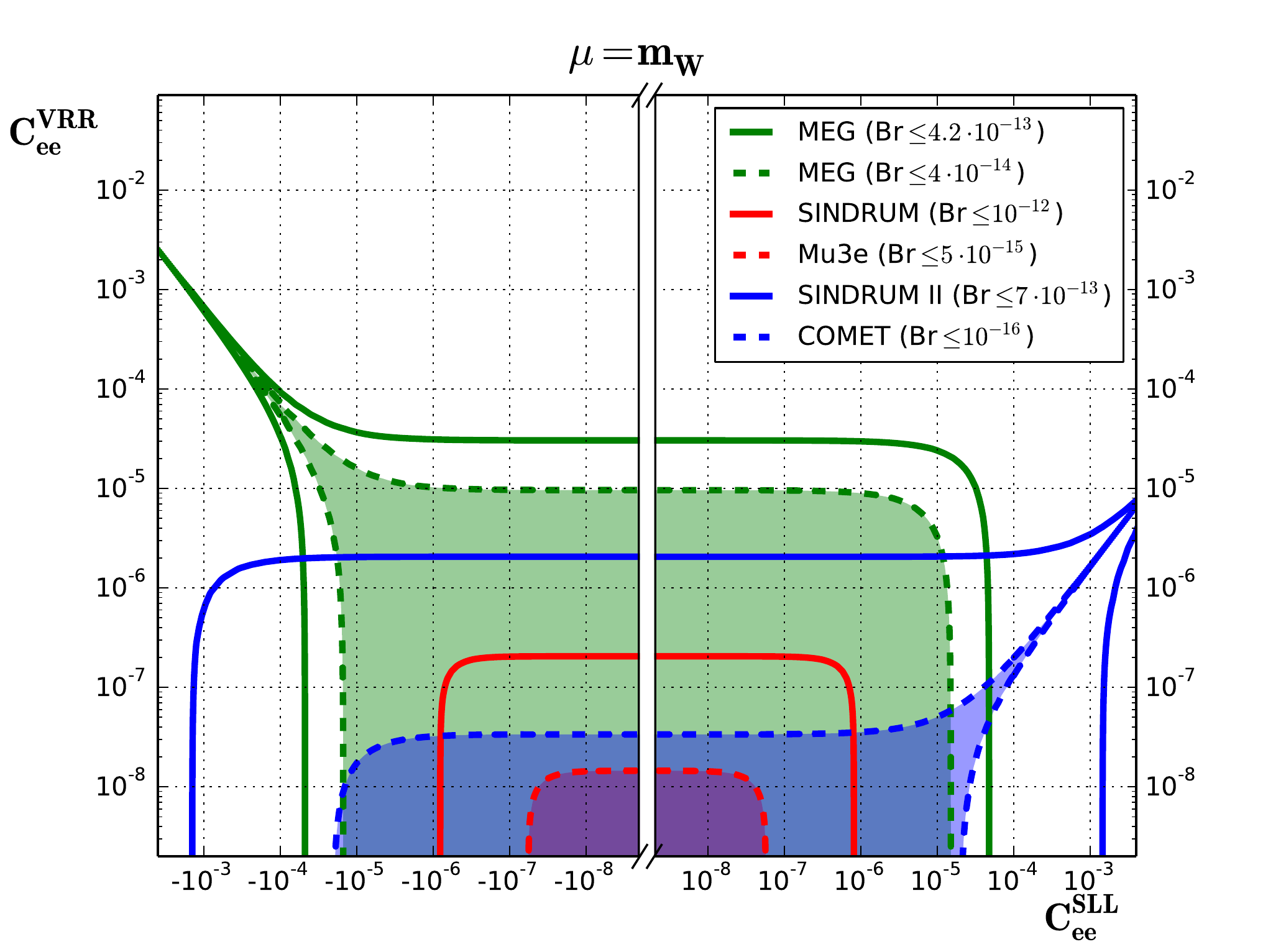} 
\end{center} 
\caption{Allowed regions in the $C^{S\,LL}_{ee}-C^{V\,RR}_{ee}$ plane
  from $\mu\to e\gamma$ (green), $\mu\to3e$ (red) and $\mu\to e$
  conversion (blue) for current (straight) and future (dashed)
  experimental limits.}
\label{figB}
\end{figure}
In Figure~\ref{figB} we show the allowed regions in the
$C^{S\,LL}_{ee}-C^{V\,RR}_{ee}$ plane (given at the scale
$m_W$). Comparing current (solid lines) and future (dashed lines)
limits on $\mu\to e\gamma$ (in green) and $\mu\to e$ conversion (in
blue) to those of $\mu\to3e$ (in red) indicate that $\mu\to3e$
experiments are most sensitive to these Wilson coefficients. It is
also interesting to note that the other experimental setups could be
blind to specific regions in parameter space where cancellations
occur, while this is never the case for $\mu\to3e$. This is mostly due
to the fact that these operators trigger $\mu\to 3e$ already at the
tree-level while they give rise to the other processes only via mixing
effects.

\begin{figure}[!th]
\begin{center}
\includegraphics[width=\textwidth]{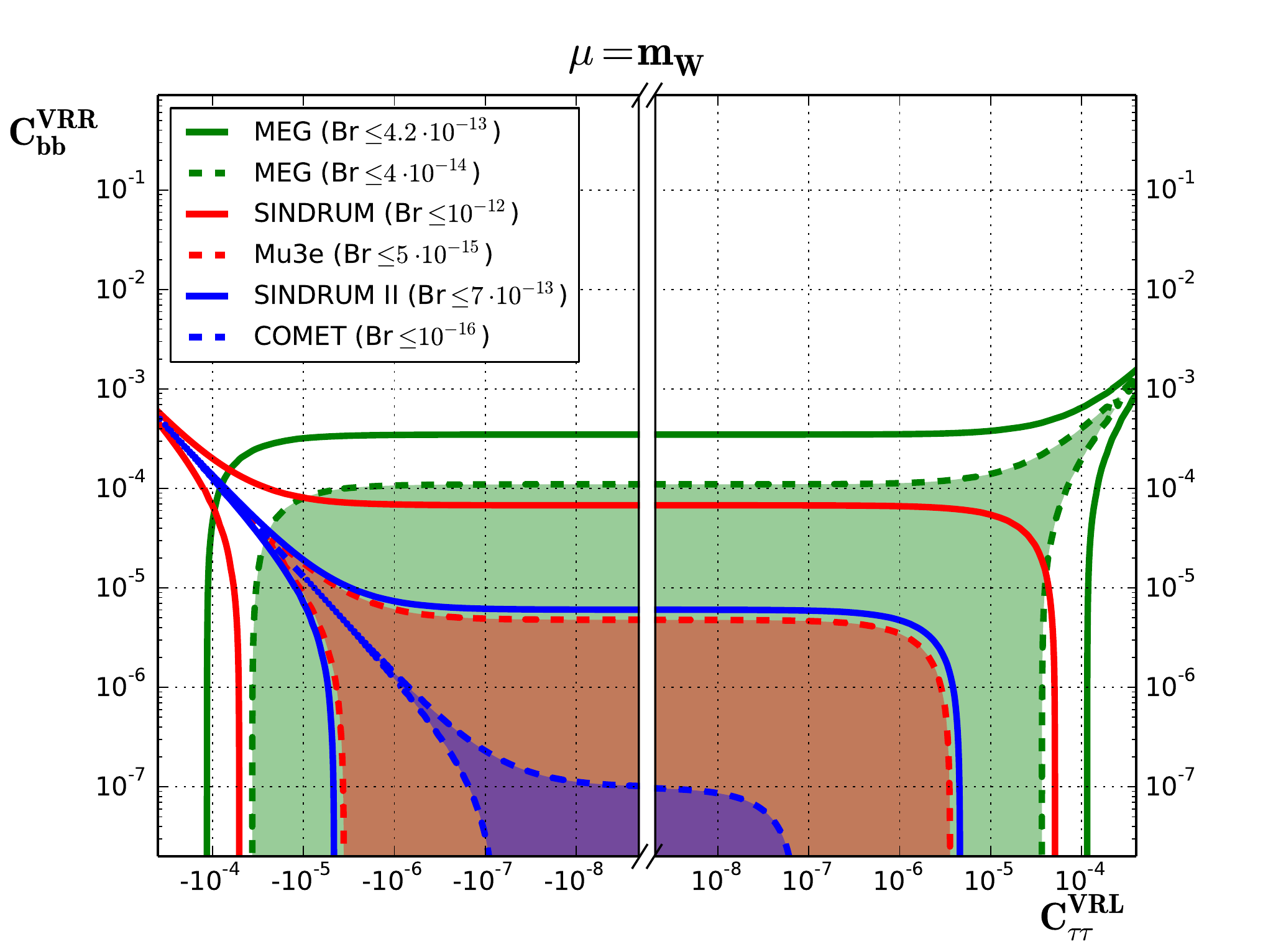}
\end{center}
\caption{Allowed regions in the
  $C^{V\,RL}_{\tau\tau}-C^{V\,RR}_{bb}$ plane from $\mu\to e\gamma$
  (green), $\mu\to 3e$ (red) and $\mu\to e$ conversion (blue) for
  current (straight) and future (dashed) experimental limits.}
\label{figA}
\end{figure}
In Figure~\ref{figA} we show an analogous plot for the Wilson
coefficients $C^{V\,RL}_{\tau\tau}$ and $C^{V\,RR}_{bb}$. In this case,
$\mu\to e$ conversion experiments display a superior capability to
probe vectorial four-fermion operators, as previously discussed. Even
the new Mu3e experiment will be just a little better than the current
$\mu\to e$ conversion limit established by SINDRUM II more than a
decade ago. However, the plot demonstrates an interesting
complementarity among various experiments: assuming that the
underlying theory produces a cancellation both in $\mu\to e$
conversion and $\mu\to 3e$, then $\mu\to e\gamma$ experiments will
provide a complementary limit, finally closing the allowed region of
the parameter space.

\begin{figure}[!th]
\begin{center}
\includegraphics[width=\textwidth]{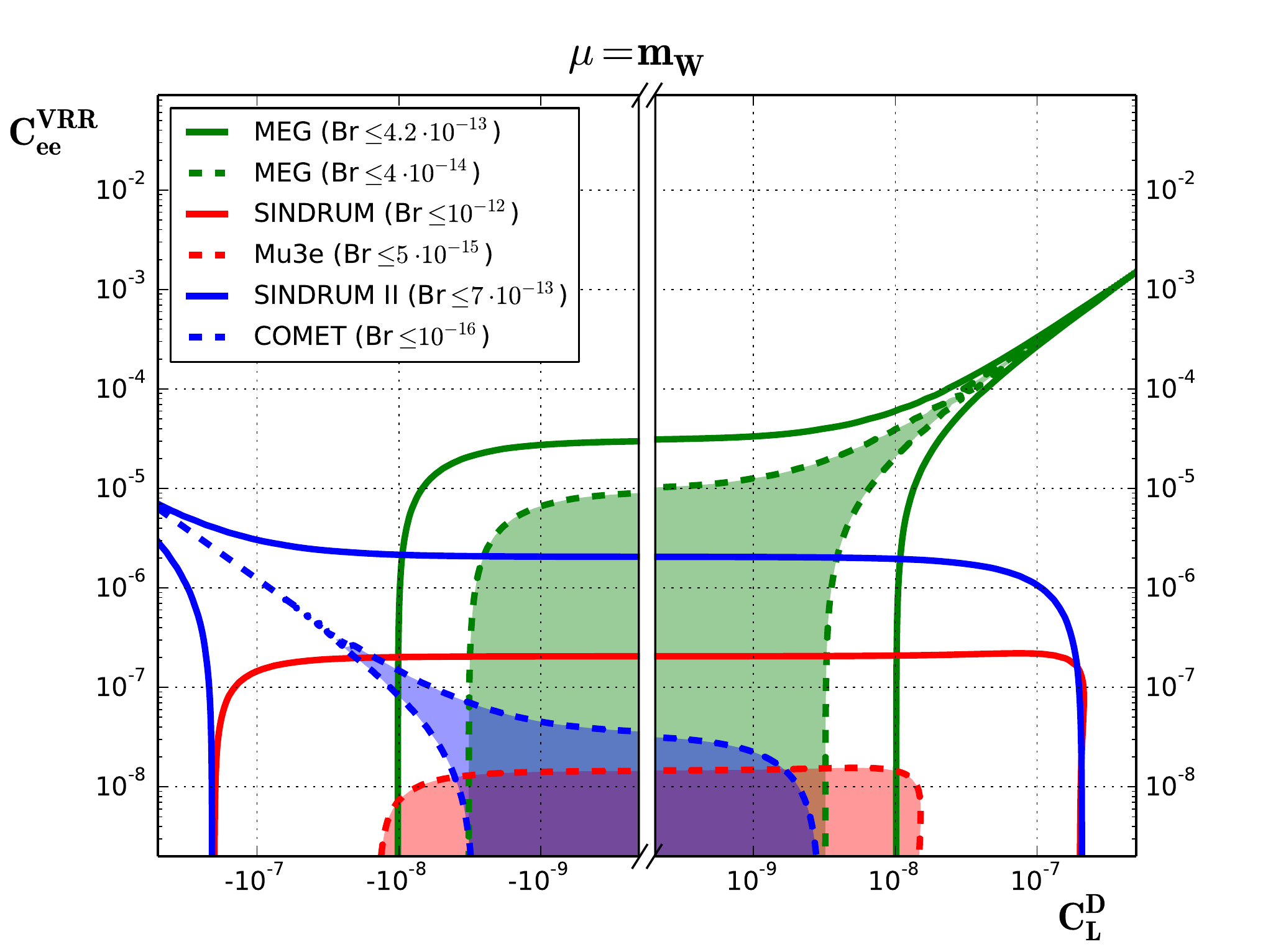}
\end{center}
\caption{Allowed regions in the $C^{D}_{L}-C^{V\,RR}_{ee}$ plane
  from $\mu\to e\gamma$ (green), $\mu\to 3e$ (red) and $\mu\to e$
  conversion (blue) for current (straight) and future (dashed)
  experimental limits.}
\label{figC}
\end{figure}
In Figure~\ref{figC}, we focus on the coefficients $C^{D}_{L}$ and
$C^{V\,RR}_{ee}$. Here, the current limit on the $C^{D}_{L}$
coefficient comes from the MEG experiment, and the future one will be
set by the $\mu\to e$ conversion. On the other hand, the past SINDRUM
and future Mu3e experiments give the most significant limits on the
coefficient $C^{V\,RR}_{ee}$. However, especially in the corners of
the parameter space where there are potential cancellations between
the contributions from $C^{D}_{L}$ and $C^{V\,RR}_{ee}$, an
interesting interplay between the observables implies that all of the
future experimental limits are useful to ensure that no blind spots in
parameter space exist.

\begin{figure}[!th]
\begin{center}
\includegraphics[width=\textwidth]{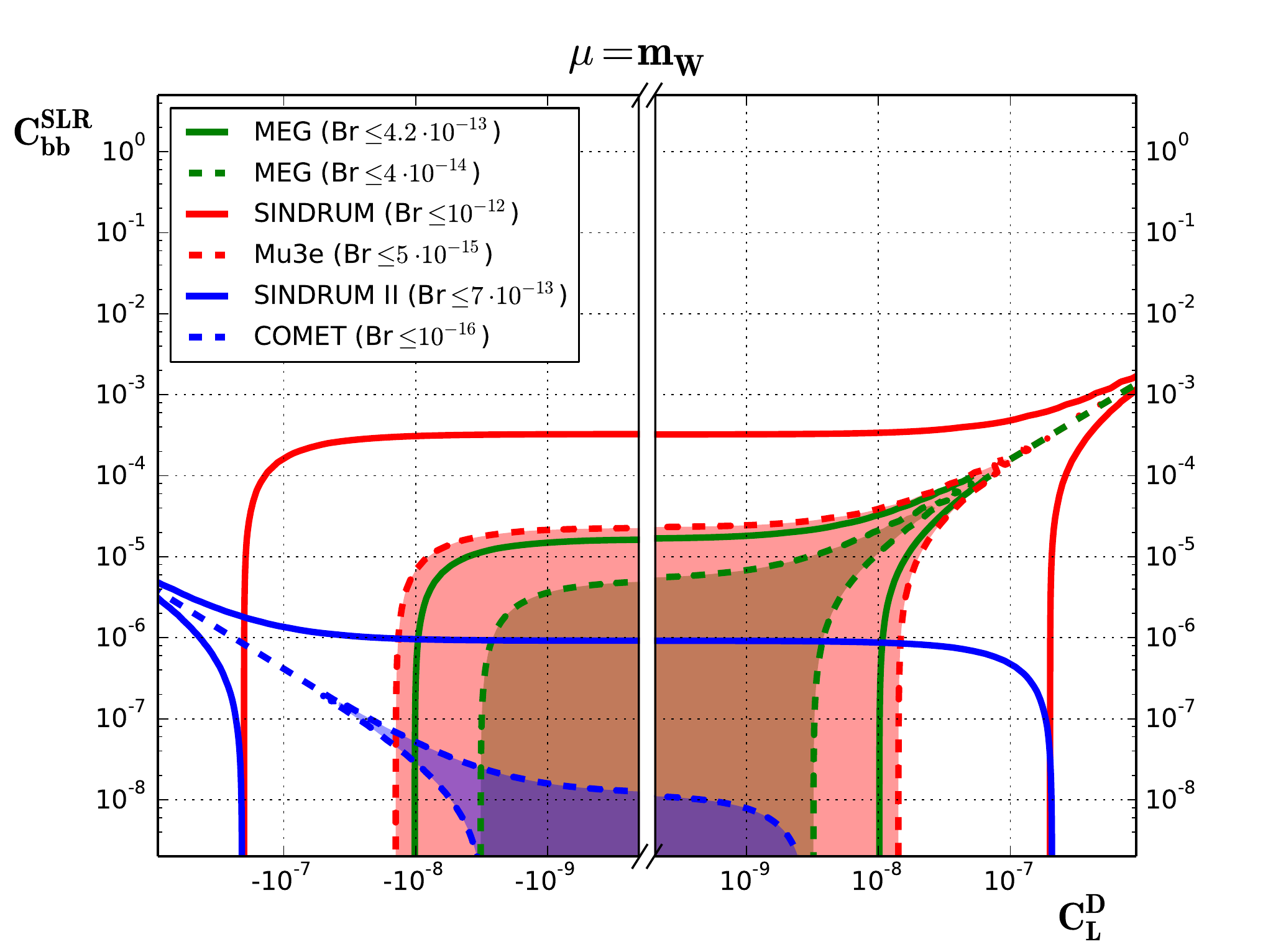}
\end{center}
\caption{Allowed regions in the $C^{D}_{L}-C^{V\,RR}_{ee}$ plane
  from $\mu\to e\gamma$ (green), $\mu\to 3e$ (red) and $\mu\to e$
  conversion (blue) for current (straight) and future (dashed)
  experimental limits.}
\label{figD}
\end{figure}
With Figure~\ref{figD} we conclude our review on correlations, by
showing the allowed regions in the $C^{D}_{L}-C^{S\:LR}_{bb}$
plane. The first information one obtains from the plot is that for
this case $\mu\to3e$ experiments are less constraining than the other
two experimental options. In the long term, $\mu\to e$ conversion will
set the best limits on each Wilson coefficient separately. However,
there is a big portion of the parameter space where a cancellation in
$\mu\to e$ conversion occurs. Results from MEG~II will play an
important role to cover this region.

Of course, our choice of combinations for the free parameters is far
from being exhaustive. However, the main message is that the interplay
between the various experiments is crucial to cover all corners of the
parameter space, also the ones in which cancellation can result in
blind spots for one or even two specific experiments.  

Obviously, it is easily possible to investigate scenarios where
several Wilson coefficients are present at the large scale. A very
efficient way to determine the impact of the experimental limits on a
particular BSM model is to obtain the Wilson coefficients at the weak
scale through matching and then use the RGE. A reasonable
approximation for the RGE can be obtained by using the numerical
evolution matrices given in Section~\ref{sec:running}. This determines
the coefficients entering Eqs.~(\ref{muegBR}), (\ref{br:meee}) and
(\ref{Gconv}) and, hence, immediately indicates whether for the chosen
parameters the model is still allowed or ruled out.

\section{Conclusions and outlook \label{sec:concl}}

In this article, we have provided RGE improved predictions for the
three $\mu\to e$ processes $\mu\to e\gamma$, $\mu\to 3e$ and $\mu\to
e$ conversion in nuclei.  Working within the effective theory valid
below the EW breaking scale, we have computed the complete one-loop
anomalous dimensions for the contributing dim-6 operators taking into
account QED and QCD effects. In addition, we have included the leading
two-loop QED effects for the mixing of vector operators into the
dipole operators and recalled the formula for the $\mu\to e\gamma$,
$\mu\to 3e$ and $\mu\to e$ conversion rates.

Our $O(\alpha_{(s)})$ RGE is renormalisation-scheme independent and
can be used for the evolution of the Wilson coefficients from any
matching scale $\Lambda \lesssim m_W$ to the scale of the experiments,
after a $O(\alpha_{(s)}^0)$ matching has been performed. If the NP
theory is realised well above the EWSB scale, this NP theory first has
to be matched to the SMEFT and the RGE within the
SMEFT~\cite{Jenkins:2013wua,Alonso:2013hga} has to be applied for the
evolution from $\Lambda$ to $m_W$. If a NP theory is realised below or
not too far above the EW scale it is sufficient to match it directly
to our Lagrangian and use the RGE discussed in this paper.

In our phenomenological analysis we have provided a numerical solution
for the RGE. In a second step, we have summarised the resulting bounds
on the Wilson coefficients (given at the scale $m_W$) under the
assumption that only one Wilson coefficient at a time is non-zero (see
Tables~\ref{tab:cc} and~\ref{tab:VA}). Afterwards, we have shown the
complementary of the three $\mu\to e$ processes by pointing out the
capability of covering regions of parameter space which would be blind
spots for a single process.

The limits presented in this paper should be interpreted in light of
the fact that they have been obtained under several simplifying
assumptions.  In particular, obtaining more accurate predictions for
the rates as a function of the Wilson coefficients is not the main aim
of including RGE contributions. More importantly, one obtains
quantitatively new effects. For example 4-fermion vector operators
with $b$, $c$ or $s$ quarks, which do not enter any of these processes
directly, mix into contributing operators resulting in stringent
constraints.  Furthermore, operators with axial-vector currents, which
do not enter $\mu\to e$ conversion at tree-level, mix into
contributing vector operators. Therefore, many more correlations among
the $\mu\to e$ processes are present once the RGE effects are taken
into account.

The future prospects for observables involving $\mu\to e$ transitions
are intriguing. MEG~II will improve the sensitivity on $\mu\to
e\gamma$ by nearly an order of magnitude, while the existing bounds on
$\mu\to 3e$ and $\mu\to e$ conversion could even improve by four
orders of magnitude. Interestingly, if $\mu\to e$ conversion managed
to improve further to $\Br \sim 10^{-18}$ it could be competitive with
$\mu\to 3e$ even for vector operators involving three electrons once
loop effects are taken into account. Furthermore, the search for
$\mu\to e$ transitions in Kaon decays like $K\to\mu e$ or $K\to\pi\mu
e$ (see~\cite{Crivellin:2016vjc} for a recent account) but also in LFV
$B$ (see for example~\cite{Crivellin:2015era,Feruglio:2016gvd}) and tau decays (see for
example~\cite{Black:2002wh,Celis:2014roa,
  Feruglio:2015yua,Pruna:2015jhf}) can give complementary
information. While, in our EFT defined below the EW scale, these
processes are completely unrelated this situation would change once
flavour symmetries are involved or if EW matching effects are
considered.


\vskip 7mm
\centerline{\bf Acknowledgements}
\vskip 2mm

\noindent AC is supported by an Ambizione Grant (PZ00P2\_154834) of
the Swiss National Science Foundation (SNSF).  GMP is supported by the
SNSF under contract 200021\_160156. GMP is thankful to Christoph
Gnendiger for the crosscheck of the anomalous dimensions of
Appendix~\ref{app:adim}.

\appendix

\section{Anomalous dimensions} \label{app:adim}

In this appendix, the running of the coefficients of the operators
listed in Eqs.~(\ref{eq:magnetic})-(\ref{Ogg}) are
presented. We use
\begin{align}
\dot{C} \equiv (4\pi)\,  \mu \dv{}{\mu}\, C
\end{align}
and $C_F=(N_c^2-1)/(2 N_c)$ with $N_c$ the number of colours.
$Q_l=-1$, $Q_u=2/3$, $Q_d=-1/3$ are the charges associated to leptons,
$u$-type and $d$-type quarks, respectively. The corresponding
equations for the chirality-flipped operators can be obtained form the
label interchange $R\leftrightarrow L$. In our computation, the
covariant derivative is defined according to the convention of {\tt
  FeynRules~v2.3}: $D_\mu\phi=\partial_\mu\phi -i g_s G^a_\mu T_a\phi$
and $D_\mu\phi=\partial_\mu\phi -i e Q_\phi A_\mu \phi$, where $\phi$
is a generic field, $G^a_\mu$ and $A_\mu$ are the gluon and photon
gauge field respectively, $T_a$ is the colour matrix and $Q_\phi$ is
the electromagnetic charge associated to the field $\phi$.

The coefficient of the dipole operator runs according
to
\begin{align}\label{rgedipole}
\dot{C}^{D}_{L}&=
16\, \alpha_{e}\, Q_l^2 C^{D}_{L}
-\frac{Q_l}{(4\pi)}\frac{m_{e}}{m_{\mu}}C^{S\;LL}_{ee}
-\frac{Q_l}{(4\pi)} C^{S\;LL}_{\mu\mu} \nonumber\\
&+ \sum_{h} \frac{8 Q_h}{(4\pi)} \frac{m_{h}}{m_{\mu}} 
   N_{c,h} \, C^{T\;LL}_{hh} \,\Theta(\mu-m_h)\nonumber\\
&- \frac{\alpha_e Q_l^3}{(4\pi)^2} \left(
    \frac{116}{9} C_{ee}^{V\;RR} 
  + \frac{116}{9} C_{\mu\mu}^{V\;RR}
  - \frac{122}{9} C_{\mu\mu}^{V\;RL}
  - \left(\frac{50}{9} + 8\, \frac{m_e}{m_\mu}\right) 
     C_{ee}^{V\;RL}\right) \nonumber\\
&- \sum_h \frac{\alpha_e}{(4\pi)^2}  \left(6 Q_h^2 Q_l
      + \frac{4 Q_h Q_l^2}{9}\right) N_{c,h} 
      \,  C_{hh}^{V\;RR} \,\Theta(\mu-m_h)\nonumber\\
&- \sum_h \frac{\alpha_e}{(4\pi)^2} \left(- 6 Q_h^2  Q_l
      + \frac{4 Q_h Q_l^2}{9}\right) N_{c,h} 
   \, C_{hh}^{V\;RL} \,\Theta(\mu-m_h)\nonumber\\
&- \sum_h \frac{\alpha_e}{(4\pi)^2} \, 4 Q_h^2 Q_l
   N_{c,h}\frac{m_{h}}{m_{\mu}} C^{S\;LR}_{hh}\,\Theta(\mu-m_h) \, ,
\end{align}
where the  sums run over  $h\in \{\tau,d,s,b,u,c\}$ with
$N_{c,\tau}=1$ and $N_{c,q}=N_c$, respectively. The terms in the last
four lines are due to two-loop contributions from vector operators (or
scalar operators that can be {Fierz-transformed} into vector operators) to
the anomalous dimension of $Q_L^D$.

The running of the whole set of vector operators is given by the
following two equations:
\begin{align}
\dot{C}^{V\;RR}_{ff}=
\frac{4\, \alpha_{e}}{3}Q_f\left(
2 Q_l \sum_{\ell=e,\mu}C^{V\;RR}_{\ell\ell}+Q_l C^{V\;RR}_{\tau\tau}
+Q_l \sum_{l}C^{V\;RL}_{ll}\right.\nonumber\\
\left.+N_c \sum_{q} Q_q\left(C^{V\;RR}_{qq}+C^{V\;RL}_{qq}\right)
+9 Q_l C^{V\;RR}_{ff}
\right),
\label{appVRR}
\end{align}
\begin{align}
\dot{C}^{V\;RL}_{ff}=
\frac{4\, \alpha_{e}}{3}Q_f \left(
2 Q_l \sum_{\ell=e,\mu}C^{V\;RR}_{\ell\ell}+Q_l C^{V\;RR}_{\tau\tau}
+Q_l \sum_{l}C^{V\;RL}_{ll}\right.\nonumber\\
\left.+N_c \sum_{q} Q_q \left(C^{V\;RR}_{qq}+C^{V\;RL}_{qq}\right)
-9 Q_l C^{V\;RL}_{ff}
\right),
\label{appVRL}
\end{align}
where $\ell\in\{e,\mu\}$, $l\in\{e,\mu,\tau\}$ and $q\in \{d,s,b,u,c\}$.

The running of the leptonic scalar and tensorial operators is
summarised by the following equations: 
\begin{align}
\dot{C}^{S\;LL}_{\ell\ell}&=
  12\,\alpha_{e}\,Q_l^2 C^{S\;LL}_{\ell\ell} \qquad \mbox{for\ }   \ell\in\{e,\mu\},
\\[8pt]
\dot{C}^{S\;LL}_{\tau\tau}&=
-12\,\alpha_{e}\,Q_l^2 
\left(
C^{S\;LL}_{\tau\tau}+8C^{T\;LL}_{\tau\tau}
\right),
\\[8pt]
\dot{C}^{S\;LR}_{\tau\tau}&=
-12\,\alpha_{e}\,Q_l^2 C^{S\;LR}_{\tau\tau},
\\[8pt]
\dot{C}^{T\;LL}_{\tau\tau}&=
-2\,\alpha_{e}\,Q_l^2 
\left(
C^{S\;LL}_{\tau\tau}-2C^{T\;LL}_{\tau\tau}
\right).
\end{align}
The running of the scalar and tensorial quark operators is given by 
\begin{align}
\dot{C}^{S\;LL}_{qq}&=
\left(-6\,\alpha_{e}\left(Q_l^2+Q_q^2\right)
-6 C_F\,\alpha_s \right)C^{S\;LL}_{qq}
-96\,\alpha_{e}\,Q_l Q_q C^{T\;LL}_{qq}+m_q k_{gg}C_{gg}^L,
\\[8pt]
\dot{C}^{S\;LR}_{qq}&=
\left(-6\,\alpha_{e}\left(Q_l^2+Q_q^2\right)
-6 C_F\,\alpha_s\right)C^{S\;LR}_{qq}+m_q k_{gg}C_{gg}^L,
\\[8pt]
\dot{C}^{T\;LL}_{qq}&=
-2\,\alpha_{e}\,Q_l Q_qC^{S\;LL}_{qq}
+\left(2\,\alpha_{e}\left(Q_l^2+Q_q^2\right)+2C_F\,\alpha_s\right)C^{T\;LL}_{qq},
\end{align}
where
\begin{align}
k_{gg}= 96\pi\,  C_F\, \alpha^2_s m_\mu G_F.
\end{align}
Finally, the gluon operator is defined in \Eqn{Ogg}  such that its
Wilson coefficient does not run at one loop,
$\dot{C}_{gg}^L=\dot{C}_{gg}^R=0$~\cite{Spiridonov:1984br}.


\end{document}